\documentclass[utf8]{frontiersSCNS} 

\setcitestyle{square} 
\usepackage{url,hyperref,lineno,microtype,subcaption}
\usepackage[onehalfspacing]{setspace}

\usepackage{graphicx}
\usepackage[acronym]{glossaries}
\usepackage{float}
\usepackage[section]{placeins}
\usepackage{varwidth}
\usepackage{xcolor}
\usepackage{hyperref}

\usepackage[shortlabels]{enumitem}
\usepackage{mathtools}

\graphicspath{{./figures/}{../jpeg/}}
\DeclareGraphicsExtensions{.pdf,.jpeg,.png,.svg}

\def\firstAuthorLast{Engebraaten {et~al.}} 
\def\Authors{Sondre A. Engebraaten\,$^{1,2,*}$, Jonas Moen\,$^{1,2}$ Oleg A. Yakimenko\,$^{3}$ \\and Kyrre Glette\,$^{1,2}$}

\def\Address{$^{1}$ University of Oslo, Norway\\
$^{2}$ Norwegian Defence Research Establishment, Kjeller, Norway\\
$^{3}$ Naval Postgraduate School, Monterey, CA, USA }
\def\corrAuthor{Sondre A. Engebraaten,\\Instituttveien 20, 2007 Kjeller, Norway}

\def\corrEmail{sondre.engebraten@ffi.no}

\newacronym{cots}{COTS}{Commercial-Of-The-Shelf}
\newacronym{pdoa}{PDOA}{Power Differential Of Arrival}
\newacronym{rss}{RSS}{Received Signal Strength}
\newacronym{uav}{UAV}{Unmanned Aerial Vehicle}
\newacronym{usv}{USV}{Unmanned Surface Vehicle}
\newacronym{rf}{RF}{Radio Frequency}

\begin{document}
\onecolumn
\firstpage{1}

\title[A Framework for Multi-Function Swarms]{A Framework for Automatic Behavior Generation in Multi-Function Swarms}

\author[\firstAuthorLast ]{\Authors} 
\address{} 
\correspondance{} 

\extraAuth{}

\maketitle

\begin{abstract}

Multi-function swarms are swarms that solve multiple tasks at once. For example, a quadcopter swarm could be tasked with exploring an area of interest while simultaneously functioning as ad-hoc relays. With this type of multi-function comes the challenge of handling potentially conflicting requirements simultaneously. Using the Quality-Diversity algorithm MAP-elites in combination with a suitable controller structure, a framework for automatic behavior generation in multi-function swarms is proposed. The framework is tested on a scenario with three simultaneous tasks: exploration, communication network creation and geolocation of \gls{rf} emitters. 
A repertoire is evolved, consisting of a wide range of controllers, or behavior primitives, with different characteristics and trade-offs in the different tasks. This repertoire would enable the swarm to transition between behavior trade-offs online, according to the situational requirements. Furthermore, the effect of noise on the behavior characteristics in MAP-elites is investigated. A moderate number of re-evaluations is found to increase the robustness while keeping the computational requirements relatively low. A few selected controllers are examined, and the dynamics of transitioning between these controllers are explored. Finally, the study develops a methodology for analyzing the makeup of the resulting controllers. This is done through a parameter variation study where the importance of individual inputs to the swarm controllers is assessed and analyzed.



%


\end{abstract}

\section{Introduction}

\begin{figure}[h]
\centering
\includegraphics[width=0.8\textwidth]{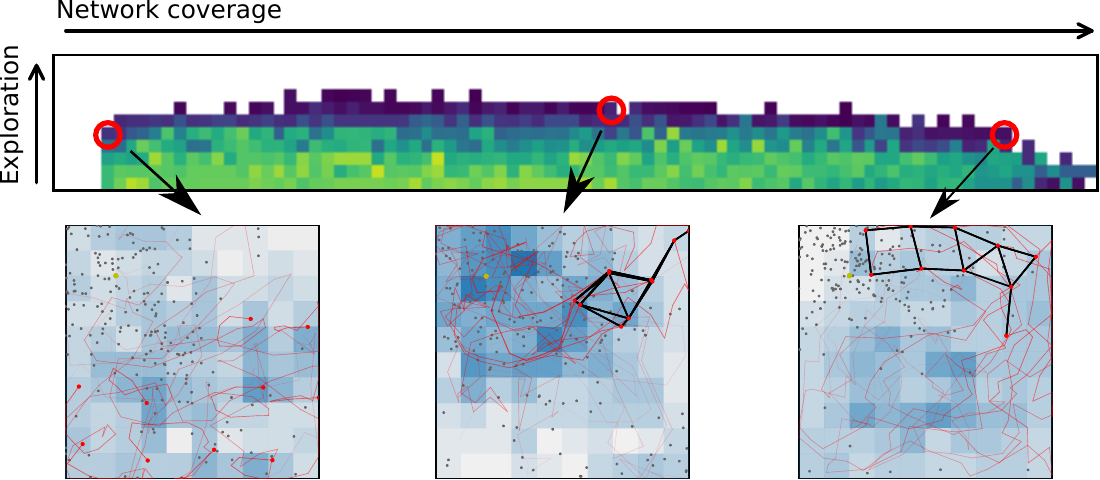}
\caption{Evolving repertoires of swarm behaviors allows a user to adapt the behavior of the swarm by simply selecting a new behavior from the repertoire. The upper figure shows part of a repertoire where a few selected controllers are highlighted.}
\label{fig:concept}
\end{figure}

Typical applications for swarms is tasks that are either too big or complex for single agents to do well. Although it might be possible to imagine a single complex and large agent that are able to solve these tasks, this is often undesirable due to system complexity or cost. Trying to solve several tasks with optimal performance also adds to the complexity (\cite{bayindir2016review,brambilla2013swarm}), as each task may place its own requirements or demands on the system. Requirements for being a good long-distance runner are not the same as being a good sprinter. Similarly, in swarms, the requirements for being good at exploring an area are not the same as for maintaining a communication infrastructure. However, an operator that requires capacity and performance in both tasks has limited options. One way of tackling this challenge could be to launch two swarms, giving each swarm a task and operating them independently. This adds complexity to the operation and doubles the system cost. Another option is to develop a concept for a multi-function swarm. 

A multi-function swarm, or a swarm that seeks to solve multiple tasks simultaneously, is a novel concept in swarm research. This is different from multitask-assignment (\cite{brutschy2014self, meng2008self, berman2009optimized, jevtic2011distributed}) in that each agent is contributing to all tasks at the same time. It is also different from multi-modal behaviors (\cite{schrum2014evolving,schrum2012evolving,schrum2015discovering}), or behaviors to solve tasks that require multiple actions in a sequence (\cite{brutschy2014self, meng2008self, berman2009optimized, jevtic2011distributed}). A multi-function swarm tackles multiple tasks at once while retaining some performance on all the tasks simultaneously. Figure \ref{fig:concept} shows example swarm behaviors selected from a repertoire with increasing performance in the networking application from left to right.

Most swarm behaviors are defined bottom-up, where the agent-to-agent interactions are specified. For the operator or the user, the desired behavior is commonly on a macroscopic level, or considering the swarm as a whole. Deducing the required low-level rules in order to achieve a specific high-level behavior is a non-trivial problem, and subject to research (\cite{jones2018evolving, francesca2014automode}). Through the use of evolution, this paper seeks to tackle the problem of top-down automated swarm behavior generation.



Previous works show how it is possible to have robots that adapt like animals by using a repertoire of behaviors generated offline (\cite{cully2015robots}). Similar adaptation techniques should be incorporated into all robotics systems, enabling the recovery from crippling failures or simply a change in the goals of the operator (\cite{engebraaten2018evolving}). A key element of this is that adaptation must happen live, or at least in a semi-live timeframe. Evolving these behaviors online would be the optimal solution, but limited compute and real-time constraints make this infeasible.


Instead of deriving new behaviors on the fly, a swarm could be based on behavior primitives. A behavior primitive is a simple pre-computed behavior that solves some task or sub-task. In this paper, each evolved behavior is considered a behavior primitive. Each behavior represents a full solution to the multi-function behavior optimization problem, however they differ in trade-off between the applications. A swarm could easily contain many of these behavior primitives. This would allow the swarm to update its behavior on the fly, as the circumstances or requirements as given by the operator are updated. This negates the need for a full online optimization of behaviors.

A good set of behavior primitives that are easily understandable and solves common tasks goes a long way towards reducing the need for human oversight (\cite{cummings2015operator}). A common problem with scaling swarms or multi-agent systems is that the number of humans required to operate the system scales linearly with the number of agents or platforms (\cite{wang2009human,cummings2015operator}). 
This does not in a good way allow the swarm to be a capability multiplier, as it should be. 

Allowing the operator to choose from a set of predefined high-level behavior primitives stored onboard would greatly reduce the need for micromanagement and might break the linear relation between number of agents and operators. By using a Quality-Diversity method (\cite{pugh2016quality, cully2017quality}) to optimize, it is possible to use the repertoire itself to gleam some insights into the performance for individual behaviors.


In this paper the Quality-Diversity method MAP-elites (\cite{mouret2015illuminating}) is employed. MAP-elites is used to explore the search space of possible swarming behaviors. This is done to allow the operator the greatest amount of choices to adapt the system to their needs. In combination with a swarm behavior framework based on physical forces (\cite{engebraaten2018evolving}), this provides a robust and expandable system that has the required level of abstraction to be possible to transfer to real drones. 
Three tasks are explored: area surveillance, communication network creation and geolocation of \gls{rf} emitters. Each task induces new requirements onto the swarm behavior. For instance, covering an area in the surveillance task requires agents to be on the move. Coverage in the communication network application increases if agents are stationary, keeping a fixed distance to each other. It is the combination of these requirements that make this a challenging swarm use-case.


By using a well-known \gls{rf} geolocation technique based on \gls{pdoa} it is possible to give estimates of location for an unknown uncooperative \gls{rf} emitter (\cite{engebraaten2015rf}). Previously published works on \gls{pdoa} (\cite{engebraaten2017meta}) allows this method to be applied also to energy and computationally limited devices. Through extensive simulations the performance of this concurrent multi-functional swarm is evaluated, and the relevance of neighbor interactions examined. Finally, it is shown that the behaviors can indeed be used as behavioral primitives i.e. building blocks for more complex sequential behavior or as commands from an operator. 

The contributions of this paper are an extensible and rigorous framework for multi-function swarms, incorporating automated behavior generation and methods for analyzing the resulting behaviors. Data mining on the results from the evolutionary methods is essential to fully utilize all the available data, as the behaviors are simply too many to manually review. This is a major extension of previous works (\cite{engebraaten2018evolving}). Through the use of ablation, or the selective disabling or removal of parts of the controller, importance of individual sensory inputs is determined. The simulator used is updated to better reflect the reality and experiences from previous real-world tests (\cite{engebraaten2018field}). Through a combination of extensive simulations, new visualizations and a deep analysis of swarm behaviors, insights can be gained which enables more efficient use of limited computational resources for future evolutionary experiments. 

Section \ref{sec:related} presents a view on related works. Section \ref{sec:methods} present the methods, framework and simulator used in this study. Section \ref{sec:results} presents the finding and results of the simulations. Section \ref{sec:discussion} provides thoughts and views on the presented results and Section \ref{sec:conclusion} concludes the paper.

\section{Related works}
\label{sec:related}





\subsection{Controller types and methods}
Controllers in the literature for swarms vary greatly. Some propose using neural networks for control (\cite{trianni2003evolving, dorigo2004evolving, duarte2016evolution}), handwritten rules (\cite{krupke2015distributed}) or even a combination or hybrid controller structure (\cite{duarte2014hybrid}). Common for all of them is that individual robots, or agents, in some way must receive inputs from the environment or other robots. Based on this information each agent decides on what to do next. This is the basis for a decentralized swarm system and allows the swarm to be robust against single point failures. 

The controller structure in this work is an extension upon artificial potential fields (\cite{krogh1984generalized, kuntze1982methods, khatib1986real}). Artificial potential fields was originally a method of avoiding collisions for industrial robot control. Additional research allowed this method to be applied to general collision avoidance in robotics (\cite{vadakkepat2000evolutionary,park2001obstacle, lee2003artificial}). Further generalization resulted in artificial physics forces; this is known as Physicomimetics (\cite{spears2004overview}).

\subsection{Evolution of controllers}

Evolution of controllers is a common way of tackling the challenge of automated behavior generation (\cite{jones2018evolving, francesca2014automode}). Evolving a set of sequential behaviors allows agents to tackle multi-modal tasks (\cite{schrum2012evolving,schrum2014evolving,schrum2015discovering}). Similarly, evolving behavior trees allow for the evolution of controllers that can easily be understood by human operators (\cite{jones2018evolving}).

Using evolution offline, only time and available computation power limits the complexity of the problems that can be tackled. Online embodied evolution is more limited in the problem complexity, but allows for the behaviors to evolve in-vivo or in the operating environment itself (\cite{bredeche2009line, eiben2010embodied}). Using embodied evolution is a way of allowing robots to learn on the fly, but also to remove the reality gap as agents are tested in the actual environment they operate in. Combining testing of behaviors in a simulator, such as ARGoS (\cite{pinciroli2011argos,pinciroli2012argos}), with some evolution on real robots to fine-tune behaviors improves performance of evolved behaviors while retaining the benefit and speed of offline evolution (\cite{miglino1995evolving}).

Evolving a large repertoire of controllers or behaviors before the robot is deployed might allow it to recover from otherwise crippling physical and hardware faults (\cite{mouret2015illuminating,cully2015robots,cully2016evolving}). Extending on this concept it is also possible to use evolved behaviors to control complex robots as in EvoRBC (\cite{duarte2016evorbc}). When evolving controllers it is important to consider the properties of the evolutionary method chosen. In particular, Quality-Diversity methods perform better with direct encodings than indirect (\cite{tarapore2016different}), and might struggle when faced with noisy behavior characteristics or fitness metrics (\cite{justesen2019map}). Challenges with noise in traditional evolutionary optimization have been documented well (\cite{cliff1993explorations,hancock1994empirical,beyer2000evolutionary,jin2005evolutionary}). However, as the method MAP-elites is fairly new the effect of noise has not been reviewed to the same extent.

\subsection{Real-world applications of swarms}

The ability to operate not only a single \gls{uav} but multiple \glspl{uav} is beneficial (\cite{bayraktar2004experimental}). 
Multiple \glspl{uav} may offer increased performance through task allocation (\cite{how2004flight}). A controlled indoor environment allows many swarm concepts to be evaluated without the constraints and uncertainty outdoor tests might bring (\cite{lindsey2012construction, schuler2019study, kushleyev2013towards,hsieh2008decentralized,preiss2017crazyswarm}). However, finding a way to move swarms out of the labs and into the real world allows for the verification of early bio-inspired swarm behaviors (\cite{reynolds1987flocks}), and the effect of reduced communication can be investigated (\cite{hauert2011reynolds}). Flocking can also be tested on a larger scale than previously possible (\cite{vasarhelyi2018optimized}). 
 
Outdoors, the potential applications for swarms are many. Swarms could provide a communication network, as is the case in the SMAVNET project (\cite{hauert2009evolved,hauert2010communication}). Teams or swarms of \glspl{uav} may be used to survey large areas (\cite{basilico2015deploying,atten2016uav}). SWARM-BOT shows how smaller ground-based robots can work together to traverse challenging terrain (\cite{mondada2004swarm}). Pushing boundaries on what is possible, scaling a swarm still presents a challenge, but ARSENL shows that a swarm of 50 \glspl{uav} is possible in live flight experiments (\cite{chung2016live}). The main challenge, apart from logistics (\cite{mulgaonkar2012automated}) in these large experiments, is that of communication and maintaining consensus (\cite{davis2016consensus}). A new frontier for outdoor swarming might be to incorporate heterogeneous platforms with wide sets of different capabilities, further extending the number of applications for the swarm (\cite{dorigo2013swarmanoid}). Swarms of \glspl{usv} might also prove valuable in environmental monitoring of vast maritime areas (\cite{duarte2016evolution,duarte2016application}).

%
%
%
%
%

\section{Methods}
\label{sec:methods}

\subsection{A tri-functional swarm}

The proposed framework uses evolution to automatically create a large set of swarm behaviors. This set of multi-function swarm behaviors is generated based on high-level metrics that measure performance in each application. The core of the framework is the combination of evolutionary methods with a directly encoded physics-based controller, which allows the framework to produce a varied set of swarm behaviors. Three applications were chosen to evaluate the framework: area surveillance, communication network creation and geolocation of \gls{rf} emitters.  The first two were introduced in previous works (\cite{engebraaten2018evolving}), while the combination with geolocation of \gls{rf} emitters is new in this paper. 


Making a framework for a multi-function swarm requires development and research into controllers for swarm agents, adaptation of existing evolutionary methods to this task, a suitable simulator to test the proposed swarm behaviors, and realistic assumptions about the capabilities of each individual swarm agent. This section will go into additional details about each of these, starting with the structure of the controllers for each agent.

%
%
%
%
%
%
%
%
%

\subsection{Simulator setup}

These experiments employ an event-based particle simulator. Every agent is modeled as a point mass with limits on maximum velocity and acceleration. A modular architecture allows the simulator to be easily expanded with new sensor, platforms or features. Using the Celery framework for distributed computation allows for task-based parallelization. The full source code for the simulator setup used can be found at GitHub \footnote{Uploaded upon final acceptance}.

Each agent is assumed to be equipped with a radio for communication with other agents and the ground, a camera, and a simple \gls{rss} sensor. All of these are both small in size and weight and constitute a feasible sensor package for a \gls{uav}. For these experiments the agents are assumed to have a downward facing camera that can capture the ground below and look for objects of interest. To further simplify the simulation, the simulated environment does not emulate internal/external camera geometry and instead simply assumes that the area of interest can be divided into cells. Each cell is smaller than the area covered by the camera at any given time. If a sufficient number of cells are used, the method would make it likely that the entire area is covered. 

The communication radio is dual-use and acts as both the interlink for the agents (agent-to-agent communication) and the communication channel with the ground control station or other entities on the ground. In previous live-flight experiments WiFi was used (\cite{engebraaten2018field}). Newer unpublished experiments have employed a mesh radio which makes it possible to remove the need for a central WiFi router, and as such, furthers the concept of a swarm. 


Compared to previous works (\cite{engebraaten2018evolving}), a more conservative vehicle model is employed. Maximum acceleration was reduced to 1.0m/s\textsuperscript{2} and max velocity was set to 10.0m/s. This was based on the results of previous real-flight experiments (\cite{engebraaten2018field}) and is a way to compensate for the slower reaction time of the physical vehicles. 

Furthermore, it was found that the range of the controller parameters determining the behavior of the platform were in many cases too high to be readily employed on real \glspl{uav}. This led to oscillating behaviors where the time delay in the physical system could not keep up with the controller. 

\subsection{Applications in the multi-function swarm}
\label{sub:application}

\begin{figure}[h]
\centering
\begin{minipage}{60mm}
  \centering
\includegraphics[width=55mm]{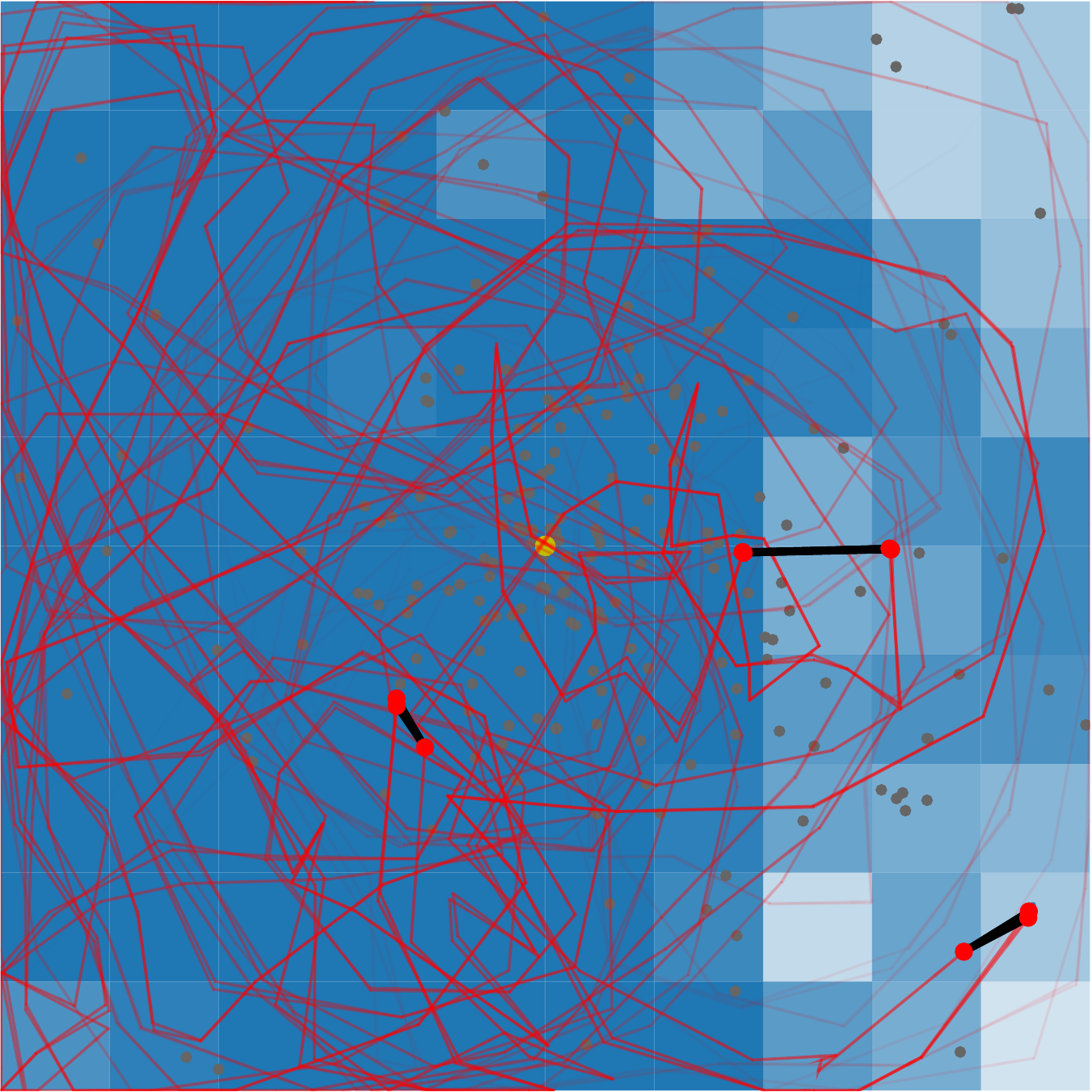}
  Exploration
\end{minipage}%
\begin{minipage}{60mm}
  \centering
\includegraphics[width=55mm]{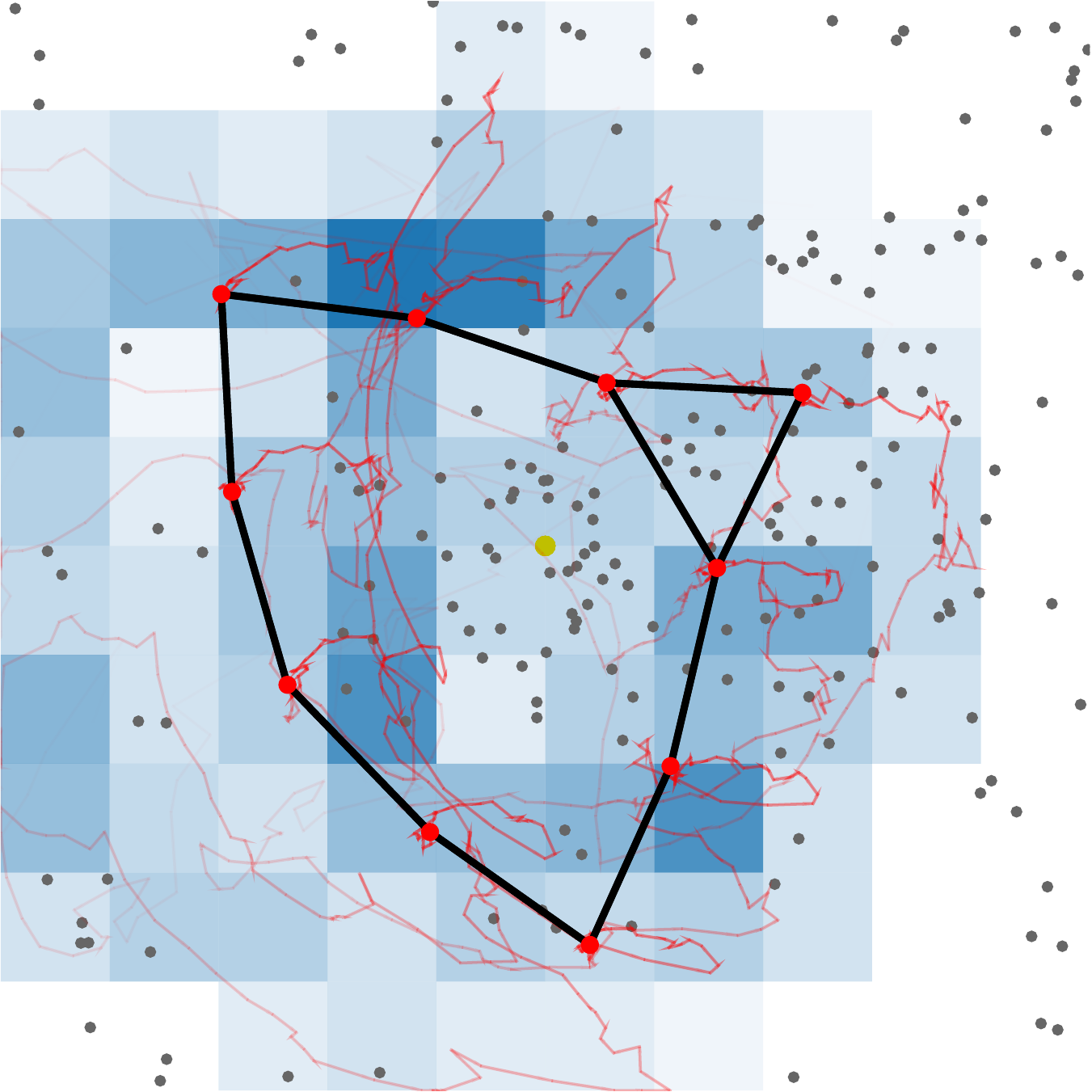}
  Network
\end{minipage}%
\begin{minipage}{60mm}
  \centering
\includegraphics[width=55mm]{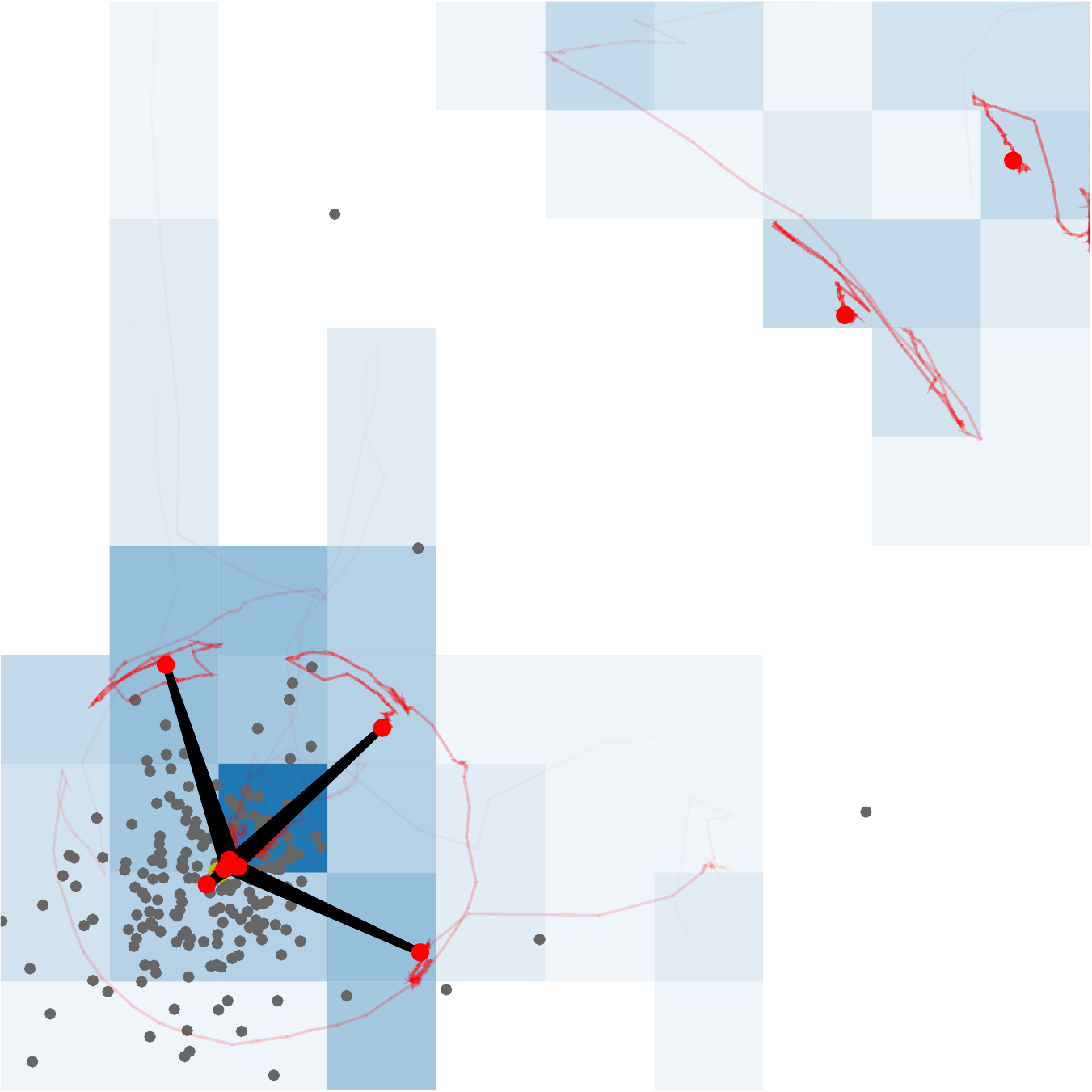}
  Geolocation
\end{minipage}
\caption{The multi-function swarm is optimized on three applications; exploration, network creation and geolocation (from left to right). Each application requires distinct behaviors for optimal performance. Red dots indicate each swarm agent.}
\label{fig:applications}
\end{figure}

A swarm of \glspl{uav} might in the future be used to provide real-time visual observations over large areas. On a conceptually high level these can be considered a potential replacement for a fixed security system, providing both better coverage, a more flexible and adaptable setup and the ability to react to new situations with ease. The downside is that today they require more maintenance and logistics, as well as more operator oversight. In this work a simplified area surveillance scenario is used as one of the applications. Each agent is equipped with a camera and tasked to explore an area. Exploration is measured by dividing the area of interest into a number of cells. The agents in the swarm seek to explore, or cover, all the cells as frequently as possible (see left part of Figure \ref{fig:applications}). The median visitation count across all cells in the area is used to measure performance in the area surveillance task.

In the absence of a common wireless infrastructure, it is also natural to imagine that the swarm must provide its own communication network in order to relay information back to an operator or a ground control station. This requires the \glspl{uav} to continuously be in range and to communicate for forwarding of data to be possible. A simplified scenario for maintaining a communication infrastructure is used as a second application for the swarm (middle part of Figure \ref{fig:applications}). Swarm behaviors are measured on the ability to maintain coverage and connectivity over a large area. The performance in the communication network task is measured by calculating the area covered by the largest connected subgroup of the swarm, given a fixed communication radius.

\Gls{pdoa} geolocation is introduced as a third application for the multi-function swarm (right part of Figure \ref{fig:applications}). Geolocation refers to trying to find the geographic position, or coordinates, of an \gls{rf} emitter based on sensor measurements. \Gls{pdoa} uses the \gls{rss}, or the received power, at multiple different points in space in order to give an estimate of the location of the emitter (right part of Figure \ref{fig:applications}). \Gls{pdoa} geolocation is a form of trilateration, or more specifically, multilateration in that the sensor readings give an indication of distance (as opposed to direction used in triangulation). A prediction of the emitter location can be made by minimizing $Q(x,y)$. $Q(x,y)$ (Eq. \ref{NLLSQxy}) is an error function that indicates the error compared to Free Space Path Loss model, given that the emitter is at coordinates $(x,y)$. $P_k$ and $P_l$ represents the \gls{rss} at positions $(x_k,y_k)$ and $(x_l,y_l)$ respectively. Previous works presented a method of providing an estimate of the location of a transmitter using significantly less resources than commonly employed estimators (\cite{engebraaten2017meta}). In this work, $Q(x,y)$ is sampled at 60 random locations and the location with the least error is used as an estimate for emitter location. This forgoes the local search used in (\cite{engebraaten2017meta}).

\begin{eqnarray}
Q(x,y) =& \sum_{k=1}^{n}\sum_{l=k}^{n} [(P_k - P_l) - 5 \alpha \log \cfrac{(x-x_l)^2 + (y-y_l)^2}{(x-x_k)^2 + (y-y_k)^2}]^2\label{NLLSQxy}
\end{eqnarray}	

Over time, multiple estimates of the emitter location are produced by the swarm. The variance of all these predictions are calculated and used as a metric for performance in the geolocation task. It is important to note that the use of \gls{pdoa} geolocation has specific requirements on sensor placement to avoid ambiguities that lead to great variance and inaccuracies in the predicted emitter locations. For more information about this, refer to previous works (\cite{engebraaten2015rf}). In most cases, variance naturally converges towards zero as the mean converges on the true mean.


\subsection{Controller framework}
\label{sub:cont_params}

Controllers for each swarm agent are based on a variant of Physicomimetics, or artificial physics (\cite{spears2004overview}). Artificial forces act between the agents and, ultimately, define the behavior of the swarm. Unlike traditional physics there is no limit on the type of forces that can act between agents. 
\begin{figure}[h]
\centering
\includegraphics[width=0.8\textwidth]{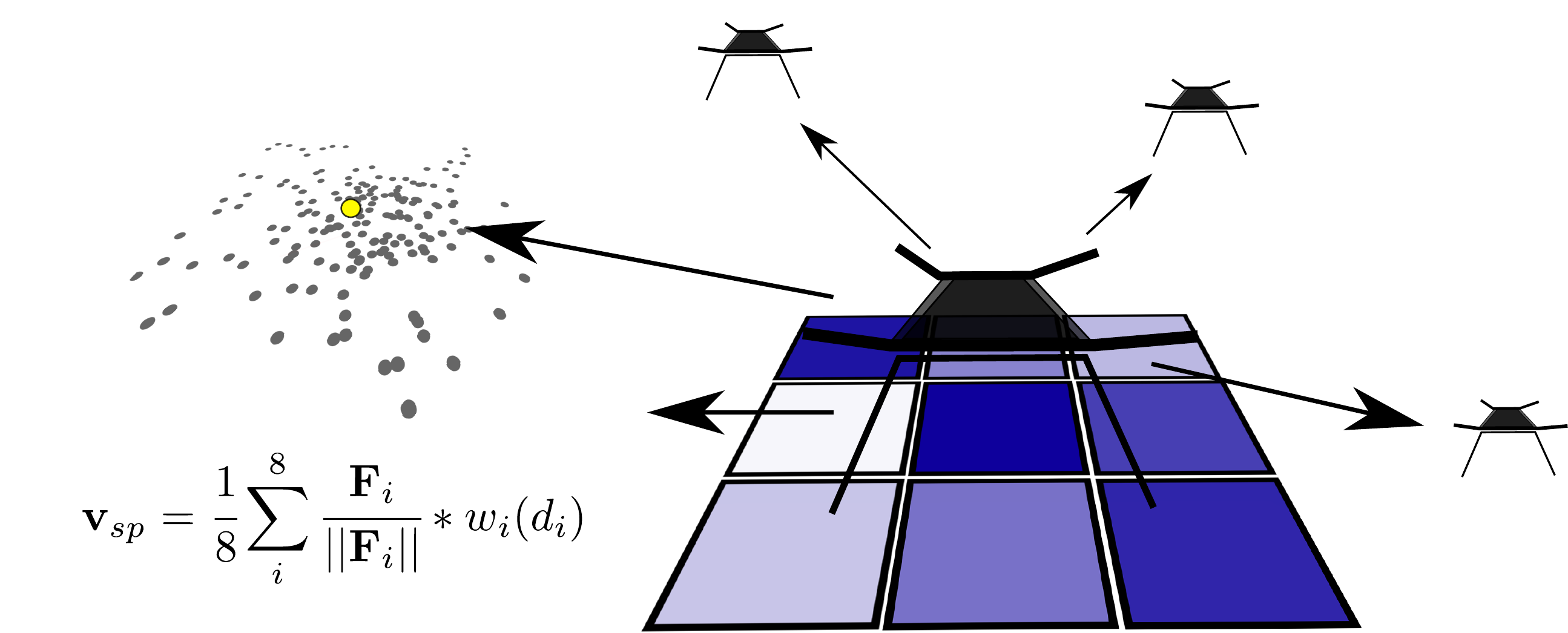}
\caption{Each agent uses the distance to the 6 nearest neighbors, the direction to the least visited surrounding square, and the direction and distance to the average of the predicted emitter location. Together, using Sigmoid-Well $w_{i}(d_i)$, they form a velocity setpoint $\mathbf{V}_{sp}$.}
\label{fig:controller}
\end{figure}
The controller (Figure \ref{fig:controller}) uses eight inputs:

\hspace{2em}
\begin{varwidth}[t]{.30\textwidth}
\begin{enumerate}[label=\textbf{F}\textsubscript{\arabic*}),labelsep=5pt]
\item Nearest neighbor
\item Second nearest neighbor
\item Third nearest neighbor
\item Fourth nearest neighbor
\end{enumerate}
\end{varwidth}
\hspace{1em}
\begin{varwidth}[t]{.5\textwidth}
\begin{enumerate}[label=\textbf{F}\textsubscript{\arabic*}),labelsep=5pt]
  \setcounter{enumi}{4}
\item Fifth nearest neighbor
\item Sixth nearest neighbor
\item Least frequently visited neighboring square
\item Average predicted emitter location
\end{enumerate}
\end{varwidth}

\begin{figure}[h]
\centering
\includegraphics[width=0.9\textwidth]{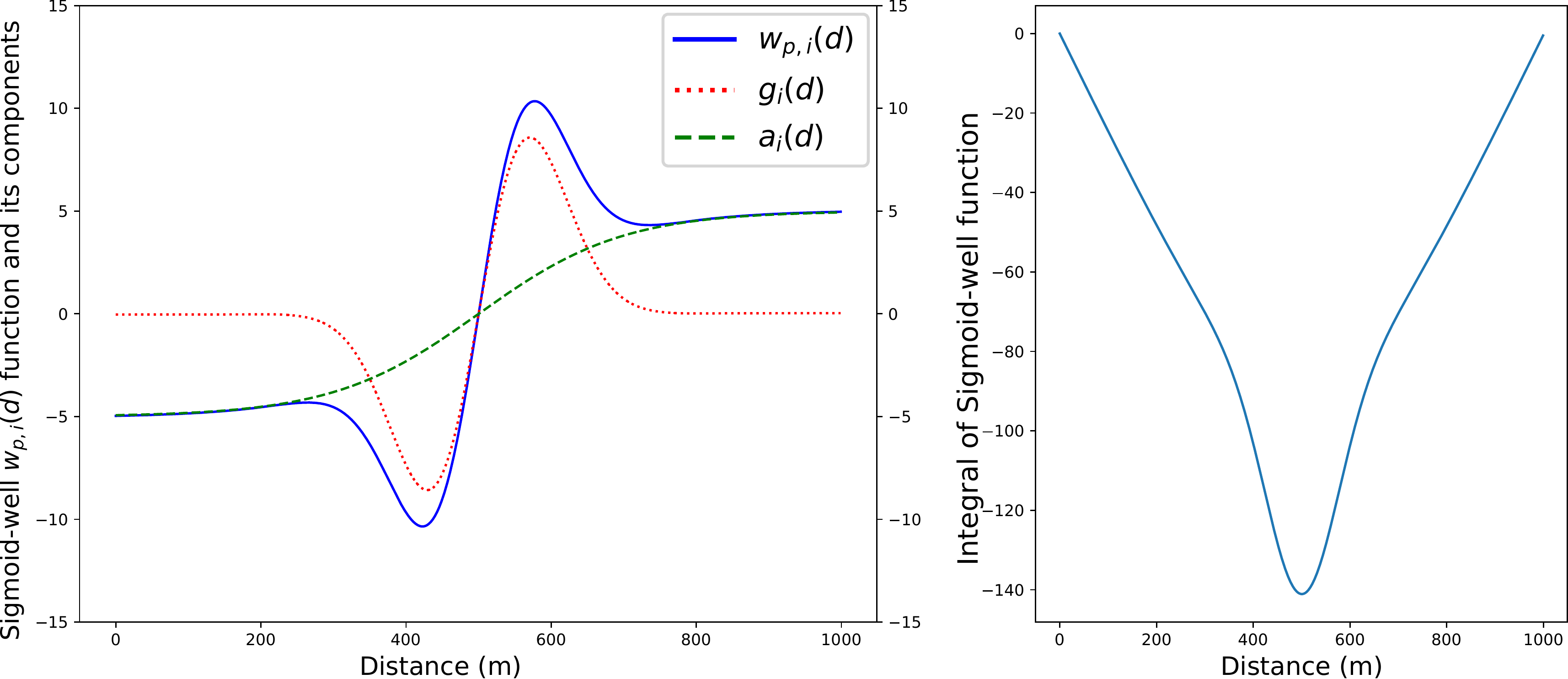}
\caption{The Sigmoid-Well function used in these experiments. Weight and scale parameters are coupled to the center parameters and spread parameters. This is because the sign of the Sigmoid-Well function changes at the center distance. Each agent minimizes the combined potential of all the contributing forces, indirectly moving to the minimum of the integral function on the right. $k_i$ is 5.0, $t_i$ is -0.1, $c_i$ is 500.0 and $\sigma_i$ is 100.}
\label{fig:sigmoidwell}
\end{figure}

In this work, the force that acts between agents is defined by the Sigmoid-Well function. This function is comprised of two parts $a_i(d_i)$ (Eq. \ref{eq:ai}) and $g_i(d_i)$ (Eq. \ref{eq:gi}). $a_i(d_i)$ is a distance dependent attraction-replusion force. $g_i(d_i)$ is the gravity well component, which can contribute with distance holding type behaviors. These functions are defined by four parameters: the $k_i$ weight, the scale parameter $t_i$ the $c_i$ center-distance and the $\sigma_i$ range parameter. The $k_i$ weight determines the strength of the attraction-repulsion force. The scale parameter $t_i$ defines the affinity towards the distance given by the center-distance $c_i$. The range parameter $\sigma_i$ can increase or decrease the effective range of the gravity well, by lengthening the transition around the center distance $c_i$. Together $a_i(d_i)$ and $g_i(d_i)$ form the Sigmoid-well function $w_{i}(d_i)$ (Eq. \ref{eq:wp}). An example of the shape of each of these components can be seen in Figure \ref{fig:sigmoidwell}, which shows how the function $w_{i}(d_i)$ can be set to enact a repulsion/attraction force, in addition to a preference for holding a distance of 500m.

\begin{eqnarray}
g_i(d_i)  &=& -t_i*2*(d_i-c_i)*e^{-(d_i - c_i)^2/\sigma_i^2} \label{eq:gi}\\
a_i(d_i)  &=&  k_i * \left( \cfrac{2}{1+e^{-(d_i - c_i)/\sigma_i}} - 1 \right) \label{eq:ai}\\
w_{i}(d_i)  &=&  a_i(d_i) + g_i(d_i) \label{eq:wp}\\
\mathbf{v}_{sp} &=& \cfrac{1}{8}\sum^8_i  \cfrac{\mathbf{F}_i}{ ||\mathbf{F}_i||} * w_{i}(||\mathbf{F}_i||) \label{eq:vsp}
\end{eqnarray}

The eight inputs are combined by scaling them with the Sigmoid-Well function $w_{i}(d_i)$ before summing the result to form a single velocity setpoint $\mathbf{V}_{sp}$ (Eq. \ref{eq:vsp}). $F_i$ is the distance delta vector from agent position to sensed object position. $\mathbf{V}_{sp}$ is calculated based on $F_i$ and $w_{i}(d_i)$.

For each input there are 4 parameters: a weight $k_i$, a scale $t_i$, a center $c_i$ and a range $\sigma_i$. With a total of 8 inputs this gives 32 parameters. The least visited neighboring square input is slightly different from the rest of the sensory inputs. This input gives only direction information to the controller and not a distance. The controller handles this by only weighting this input and not applying the distance dependent Sigmoid-Well function. This means that in practice, for each controller, only 29 parameters make an impact on the swarm behavior.

The controller inputs used in this work represent a very limited subset of everything that could be shared between agents. This is intentional, to reduce the amount of communication required. While this paper only pertains to simulation results, the goal is to fly this system outdoors. Outdoor challenges such as limited bandwidth, non-uniform antenna diagrams, loss of links, interference and link latency must be considered. When designing a swarm to work in the real world, the range and rate of communication needs to be limited. If all agents in the swarm needed information from all other agents, the system would quickly break down due to network saturation.

Compared to previous works (\cite{engebraaten2018evolving}), the ranges of the weight and scale parameters were reduced. For these experiments, the weigh parameter is limited from -2.0 to 2.0 and the scaling parameter is limited from -0.5 to 0.5. It should be noted that due to the form of the Sigmoid-Well function (\cite{engebraaten2018evolving}) the scaling parameter is stronger and has a smaller allowable range.

\subsection{Evolving controllers}



The Quality-Diversity method MAP-elites is used to evolve swarm controllers. MAP-elites seeks to explore the search space of all possible controllers by filling a number of characteristics bins spanning the search space of all controllers (\cite{mouret2015illuminating}). Variation of solutions in MAP-elites is done by mutations. In this work, mutation is performed by first selecting a parameter at random then adding a Gaussian perturbation with a mean of 0.0 and a variance that is 10\% of the range of the parameter chosen. Only a single parameter is changed per mutation. Adaptation to fill new bins might require multiple sequential mutations in order to move from one characteristics bin to another. This type of adaptation might be challenging for the MAP-elites algorithm, as it also requires that the solution outperforms existing solutions along the path required to reach the new unoccupied bins.

As part of the evolutionary process, three behavior characteristics and a fitness metric are used. As mentioned in Subsection \ref{sub:application} the characteristics or metrics are: the median visitation count across all the cells in the area of operation, the area covered by the largest connected subgroup of the swarm, and the variance in predicted locations. For the networking application each agent is assumed to have a fixed communication radius of 200m. Fitness $f$ is calculated in a deterministic manner based on the scales and weights parameters of the controller (Eq. \ref{eq:fit}). These parameters determine the magnitude of the output of the controller and as such correlate well with the motion that can be expected from the controller. In order to limit aggressiveness and reduce battery consumption, behaviors are optimized to minimize motion and maximize $f$.

\begin{eqnarray}
f &=& \cfrac{1}{||\mathbf{t}|| + ||\mathbf{k}||}\label{eq:fit}
\end{eqnarray}

\section{Experiments and results}
\label{sec:results}

A series of experiments are conducted to explore the evolutionary process itself, the viability of using the evolved swarm behaviors as behavior primitives, the effect of noise on an evolutionary process using MAP-elites and finally, the value of disabling certain controllers inputs are examined. The evolution of a single repertoire takes around 16 hours on a cluster running 132 threads; thus approximately 2112 CPU hours per repertoire. Total simulation time for the ablation study is approximately 152 064 CPU hours or 17.6 CPU years.

\subsection{Repertoire evolution}

%



%
%
%

\begin{figure}[h]
\centering
\begin{minipage}{46mm}
  \centering
\includegraphics[width=38mm]{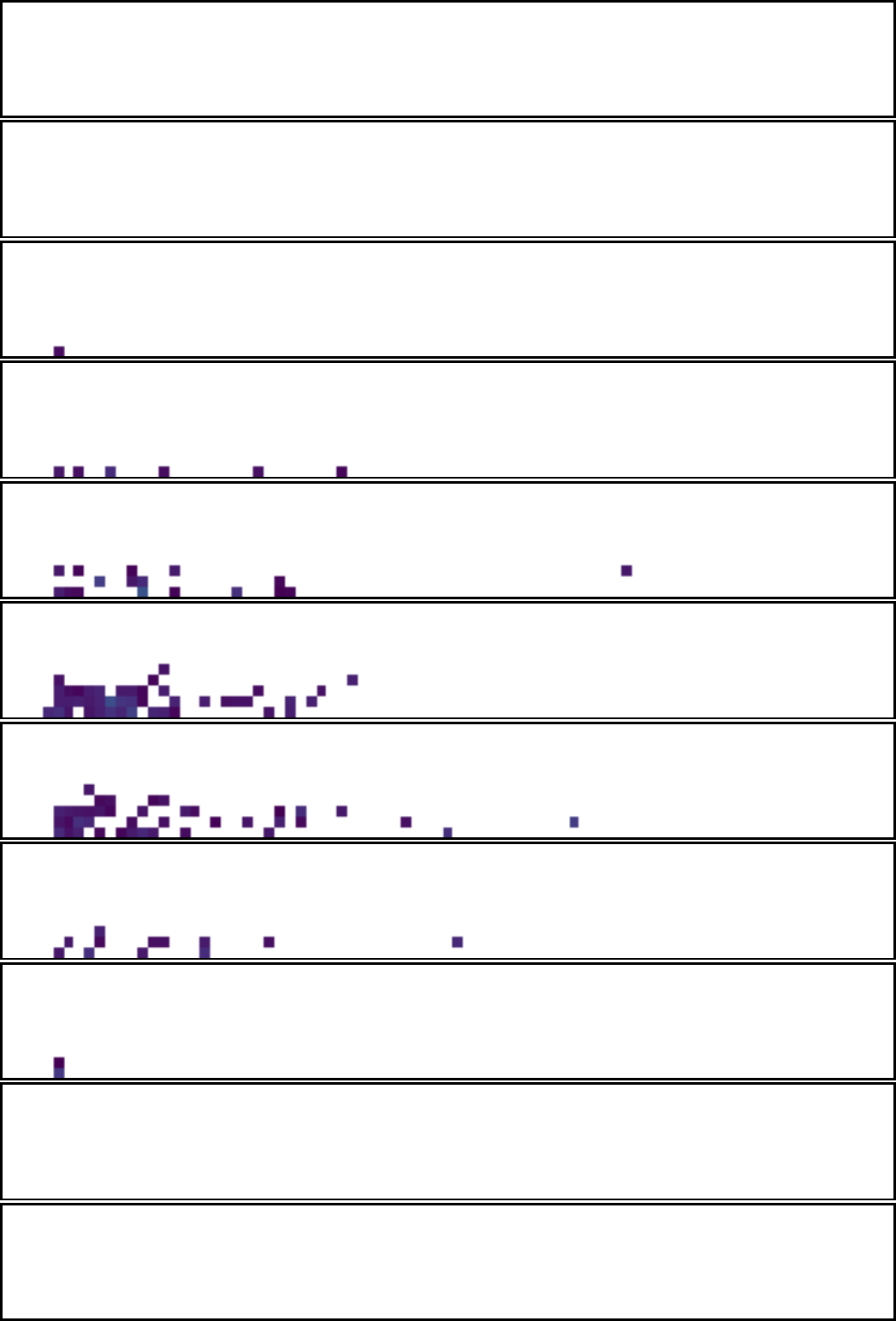}
  First generation
\includegraphics[width=38mm]{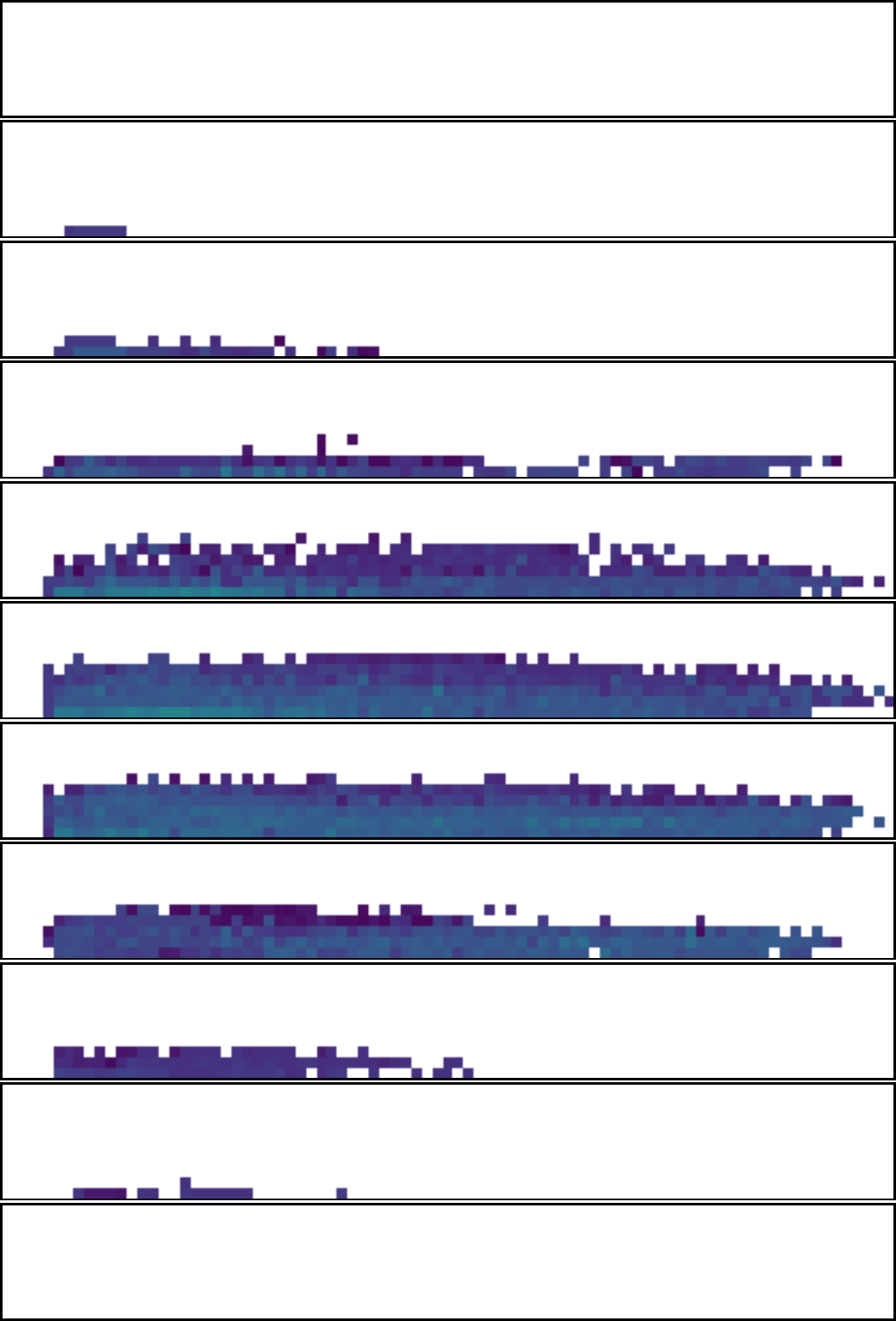}
  100th generation
\end{minipage}%
\begin{minipage}{120mm}
  \centering
\includegraphics[width=110mm]{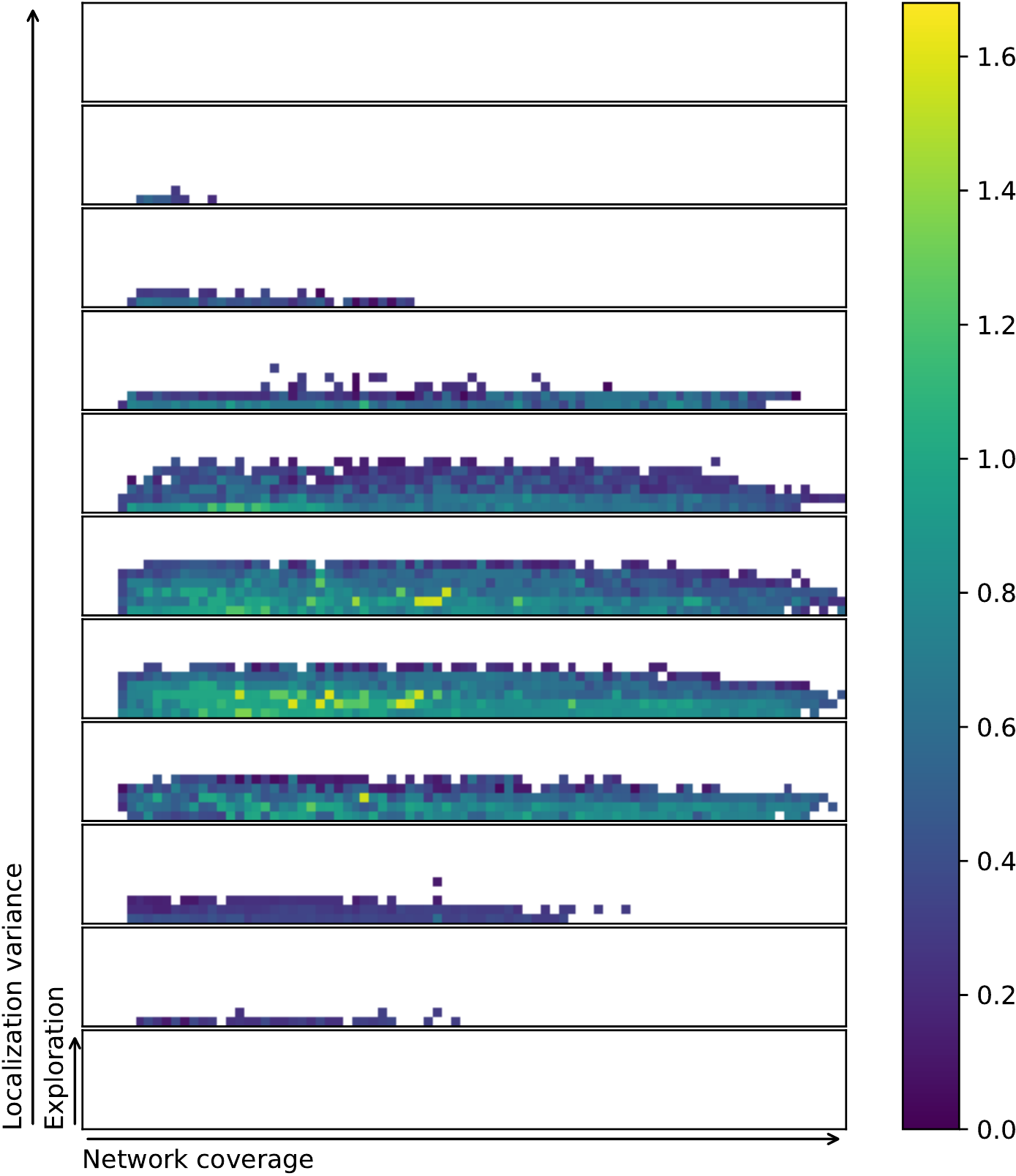}
\end{minipage}%

\caption{Progression of the evolution of one repertoire. Repertoires are visualized by slicing the three-dimensional behavior space along all three axes, this allows higher dimensions to be flattened for easier visualization. The right repertoire is the final result after 200 generations. Brighter yellow indicates better solutions, as measured by fitness.}
\label{fig:evolution_progress}
\end{figure}

All the repertoires in this work have 10 exploration bins x 100 network bins x 10 localization characteristics bins. These are filled in during 200 pseudo-generations, each evaluating and testing 200 individuals, resulting in the final repertoire. Figure \ref{fig:evolution_progress} shows the progression of evolving a single repertoire. A total of 8 independent evolutionary runs are conducted. On average, across 8 runs, the evolution resulted in 2031 solutions in each repertoire. This represents a coverage of 20.3\% with a standard deviation in number of solutions of 101.1. As can be seen in Figure \ref{fig:evolution_progress}, the first half of the evolution fills out most of the repertoire. Solutions are further optimized, and the repertoire is slightly extended during the second half of the evolutionary process.

\subsection{Swarm behaviors as behavior primitives}

\begin{figure}[h]
\centering
\includegraphics[width=0.7\textwidth]{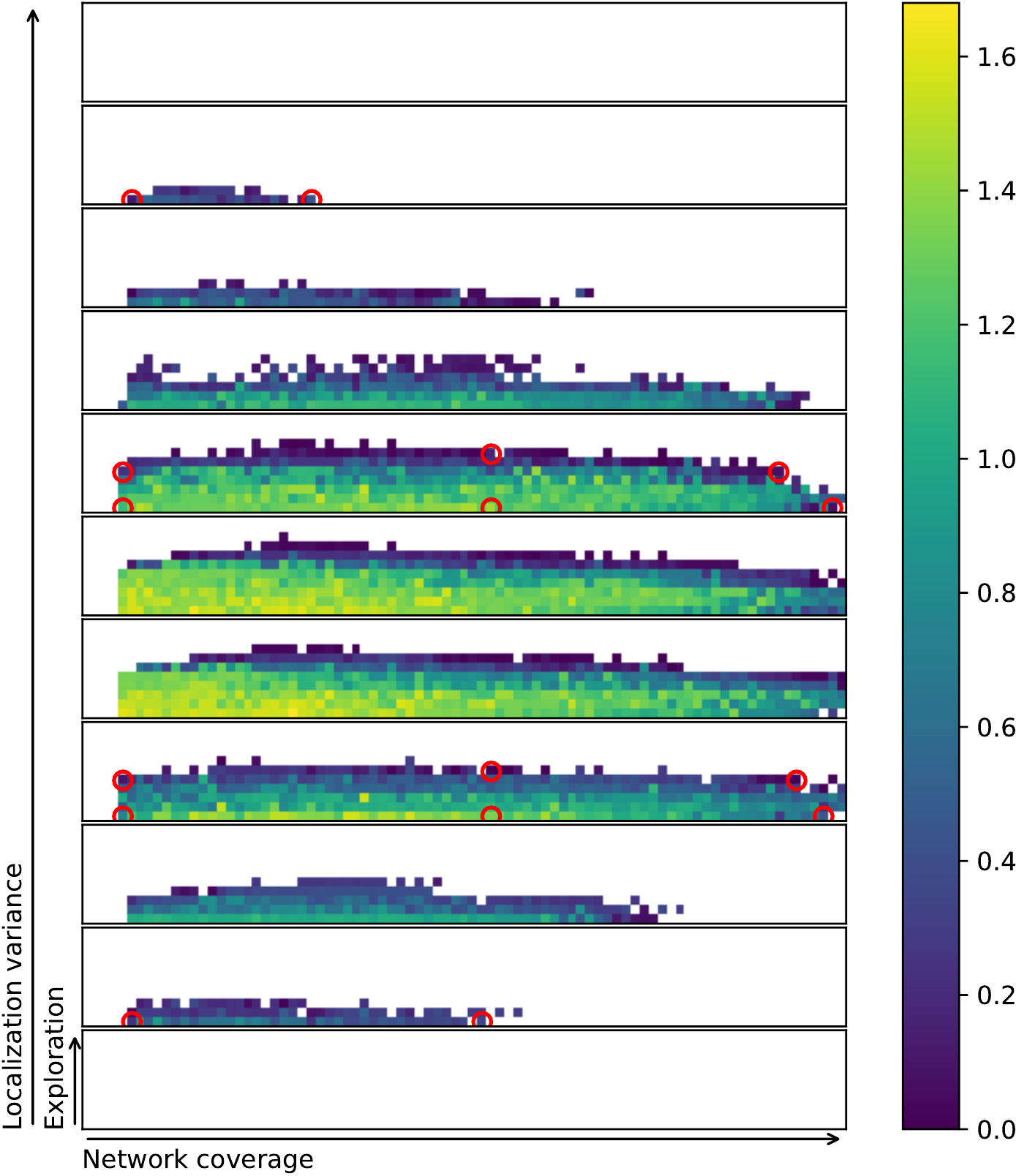}
\caption{Combined repertoire from 8 separate evolutionary runs. Each filled square represents a characteristics bin. Circled controllers are selected for a more in-depth examination. Brighter yellow indicates better solutions, as measured by fitness.}
\label{fig:mapelites_combined}
\end{figure}

\begin{figure}[h]
\centering
\begin{minipage}{37.2mm}
  \centering
\includegraphics[width=37.2mm]{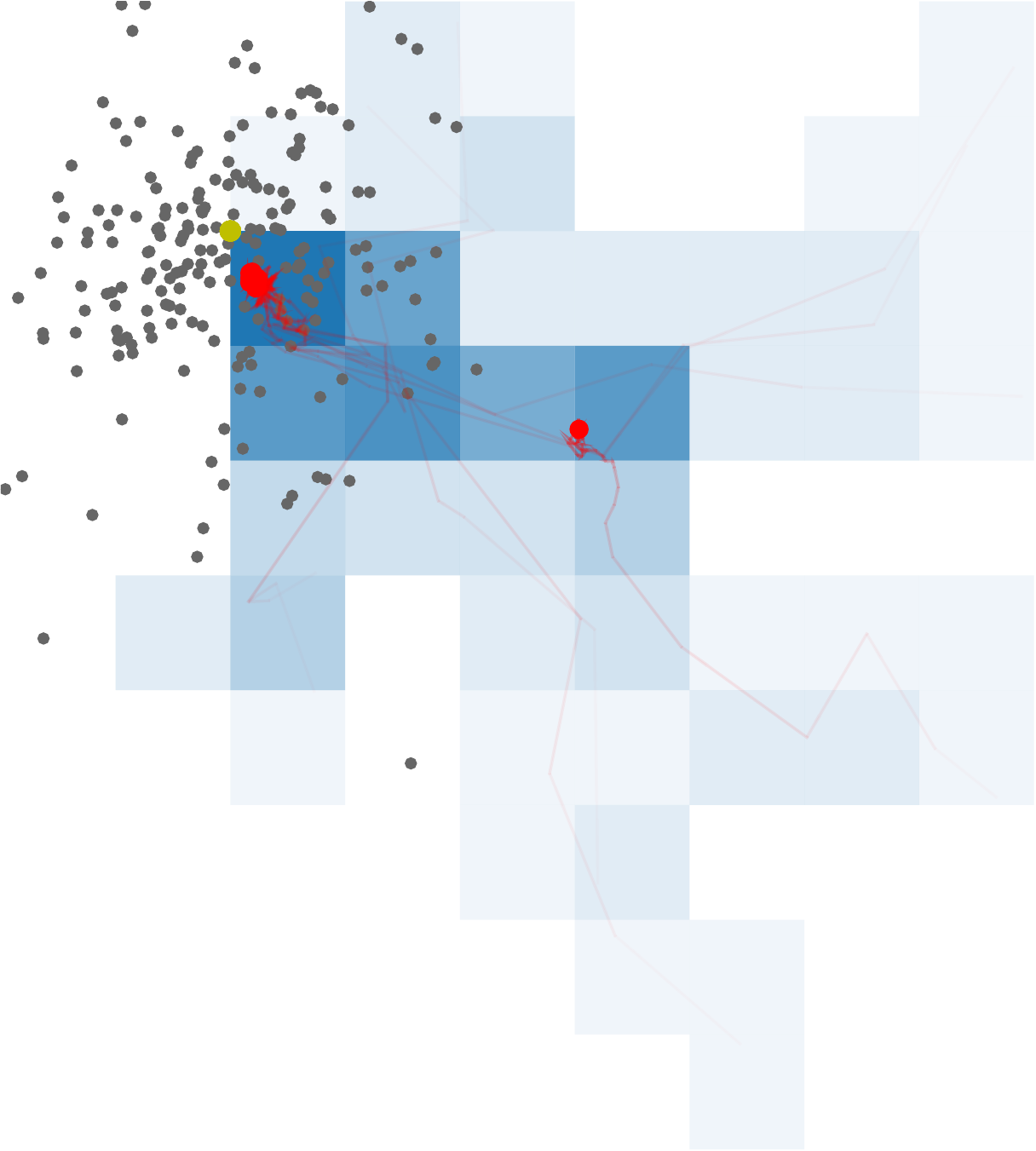}
  \href{https://www.youtube.com/watch?v=kuE3uccjc14}{LLL}
\end{minipage}
\begin{minipage}{37.2mm}
  \centering
\includegraphics[width=37.2mm]{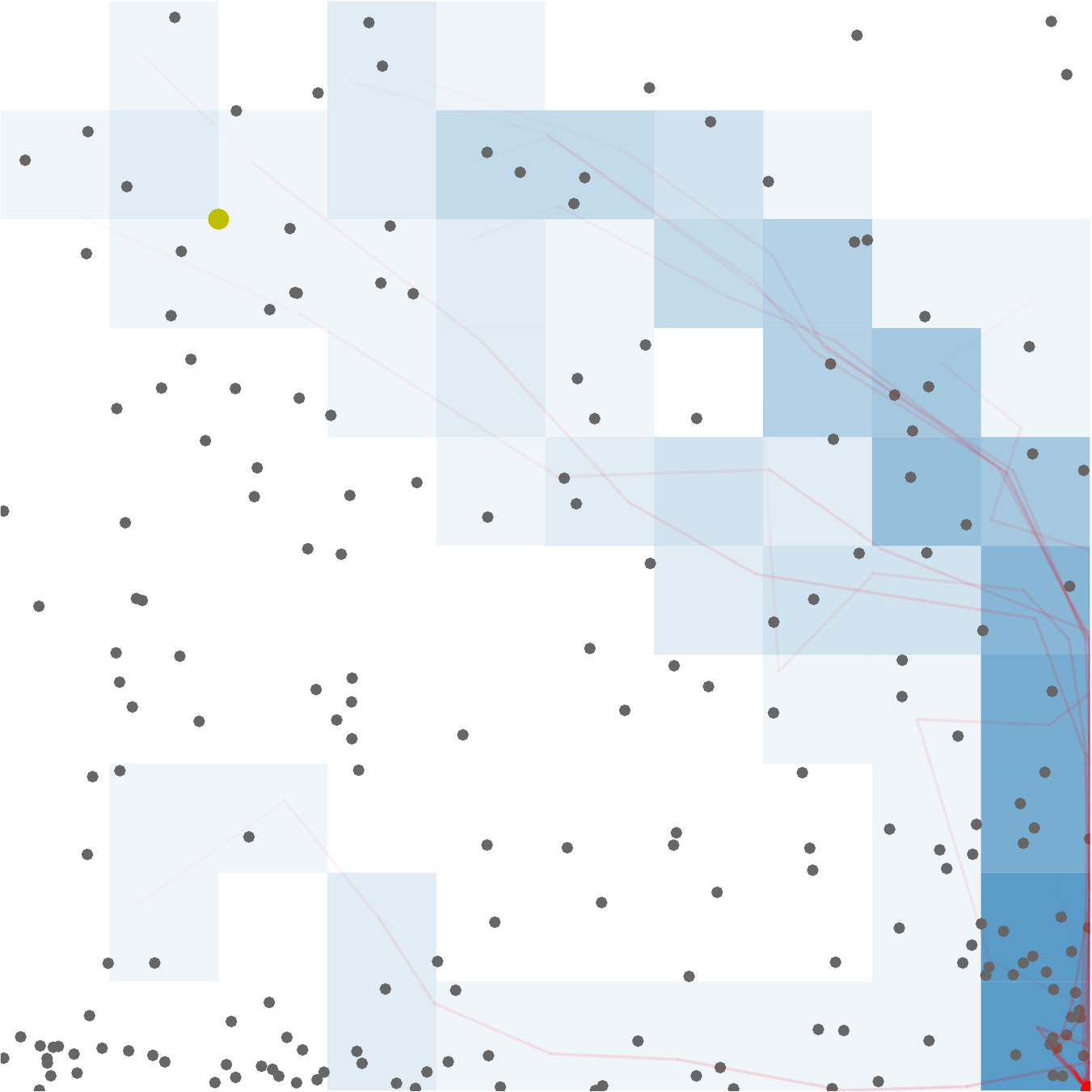}
  \href{https://www.youtube.com/watch?v=qQEtncSon48}{LLH}
\end{minipage}
\begin{minipage}{37.2mm}
  \centering
\includegraphics[width=37.2mm]{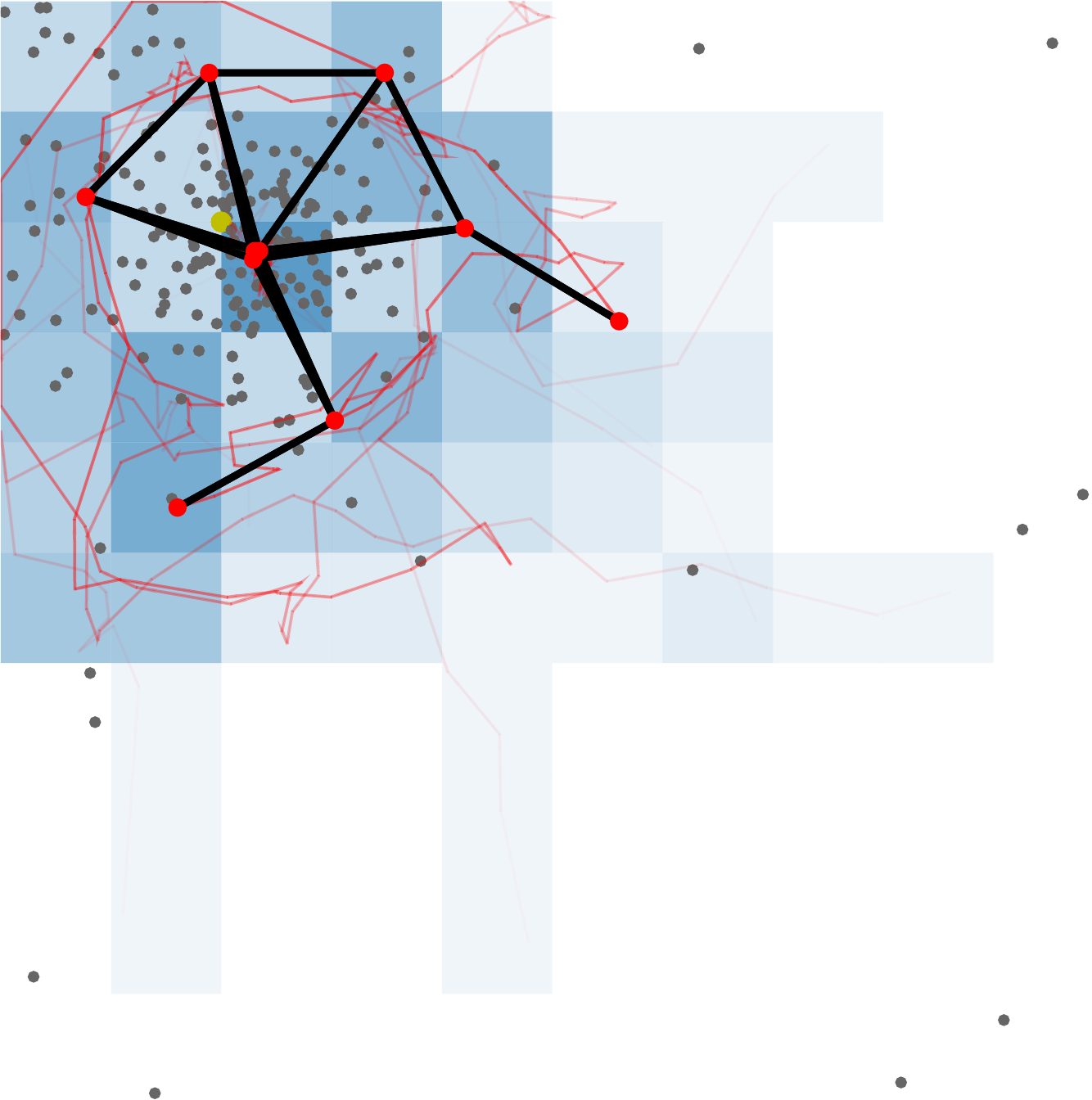}
  \href{https://www.youtube.com/watch?v=T84es0jG3yg}{LHL}
\end{minipage}
\begin{minipage}{37.2mm}
  \centering
\includegraphics[width=37.2mm]{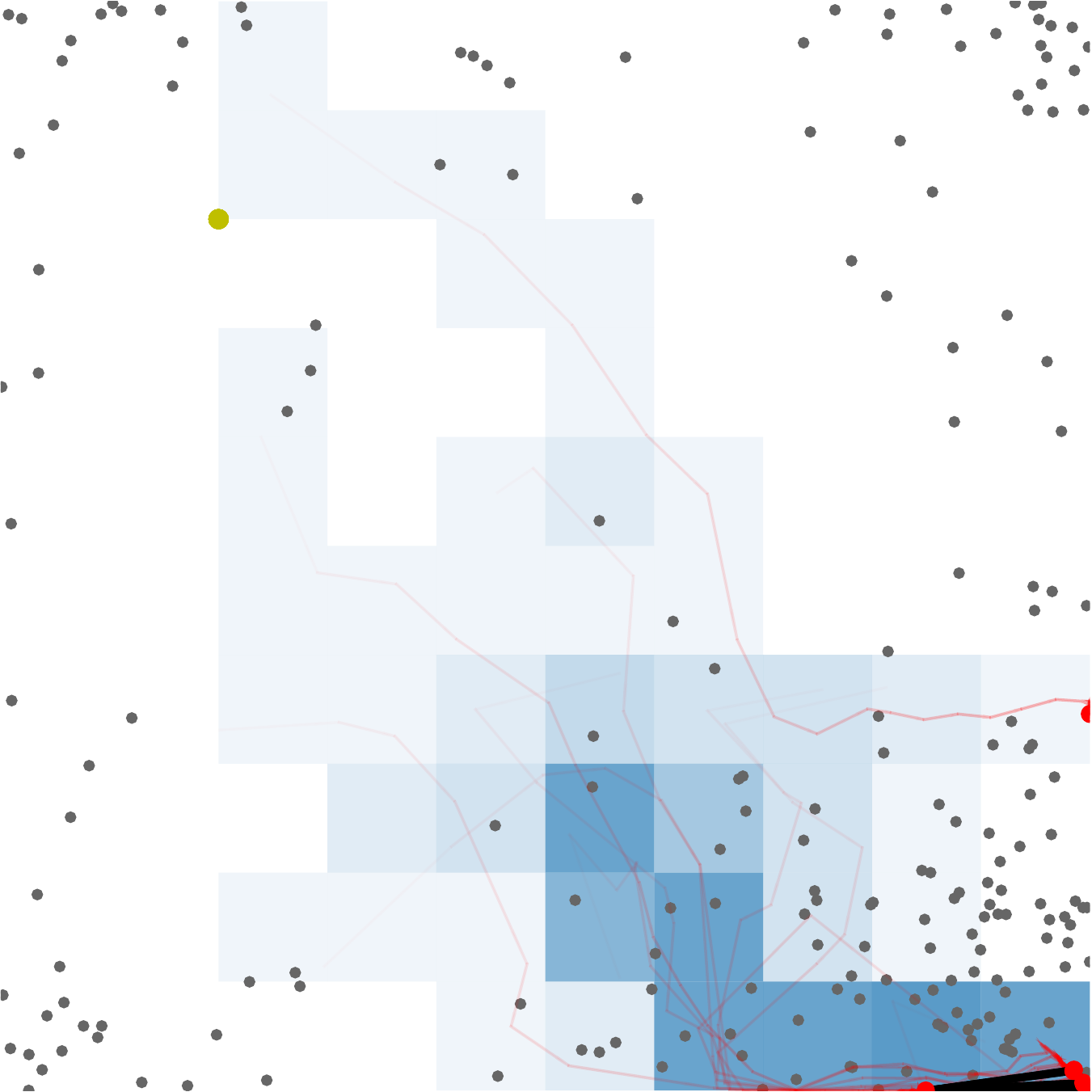}
  \href{https://www.youtube.com/watch?v=EIwJtoD7WFo}{LHH}
\end{minipage}\\
\begin{minipage}{37.2mm}
  \centering
\includegraphics[width=37.2mm]{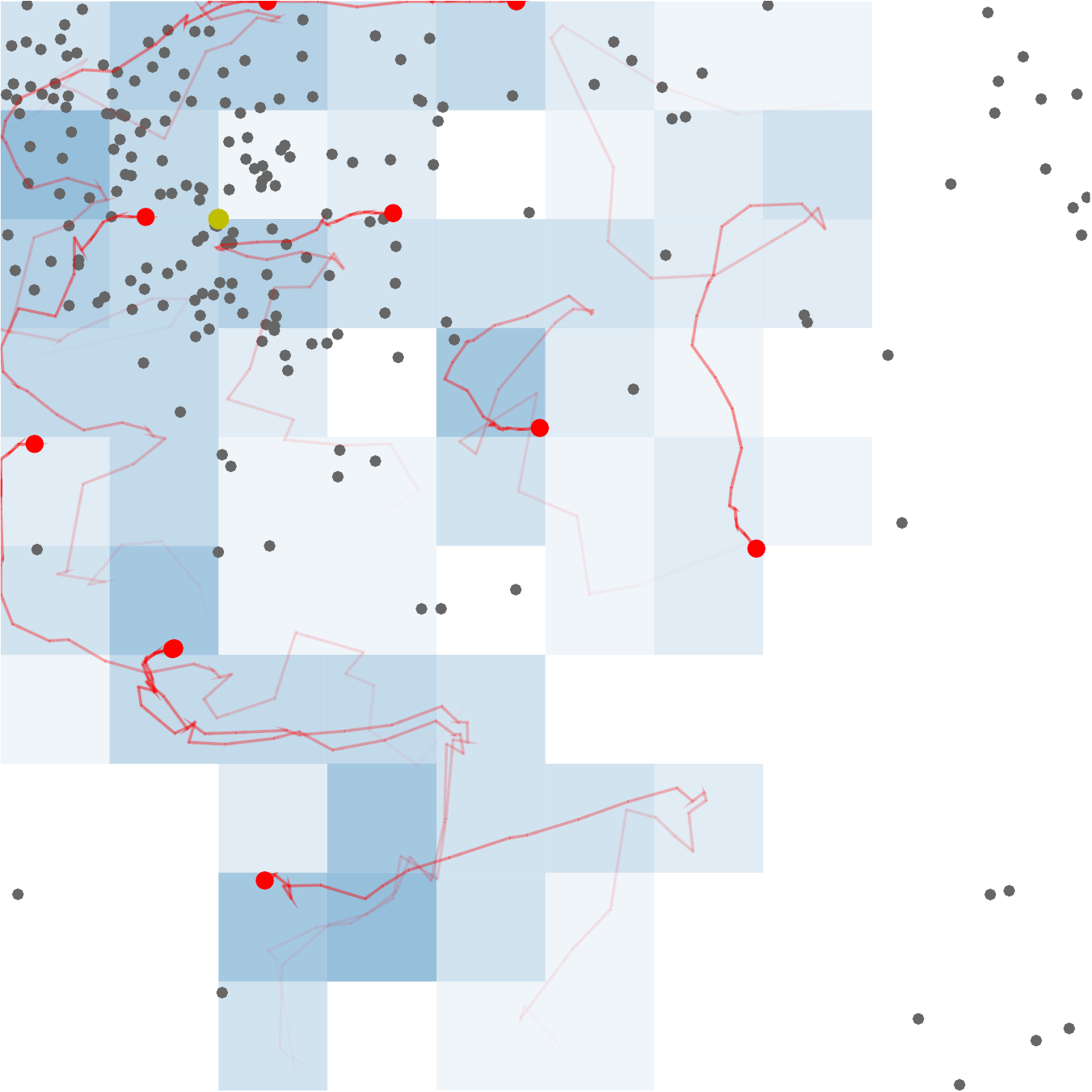}
  \href{https://www.youtube.com/watch?v=ig3Iir6lAEk}{LL4}
\end{minipage}
\begin{minipage}{37.2mm}
  \centering
\includegraphics[width=37.2mm]{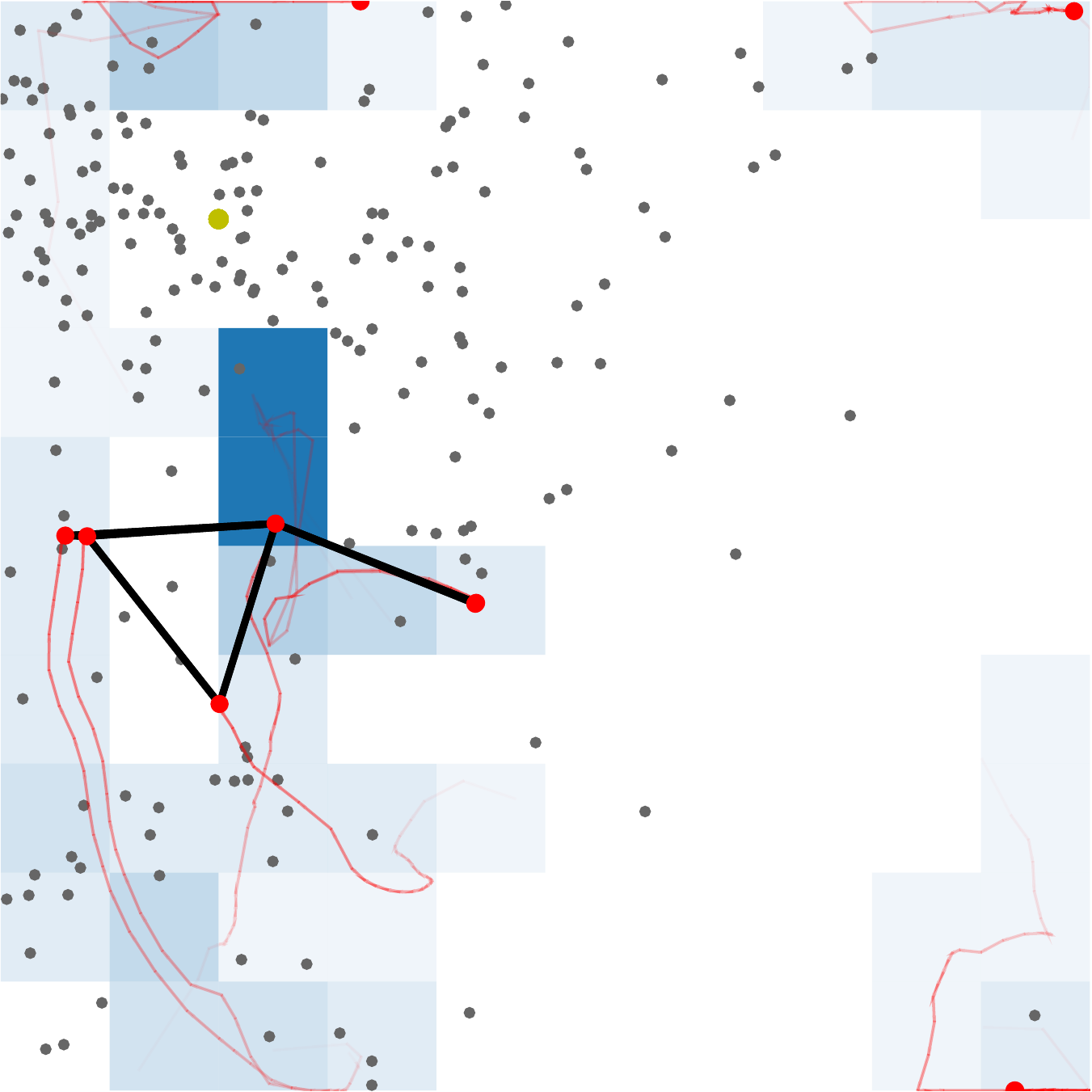}
  \href{https://www.youtube.com/watch?v=VxzXtcQBw0s}{LM4}
\end{minipage}
\begin{minipage}{37.2mm}
  \centering
\includegraphics[width=37.2mm]{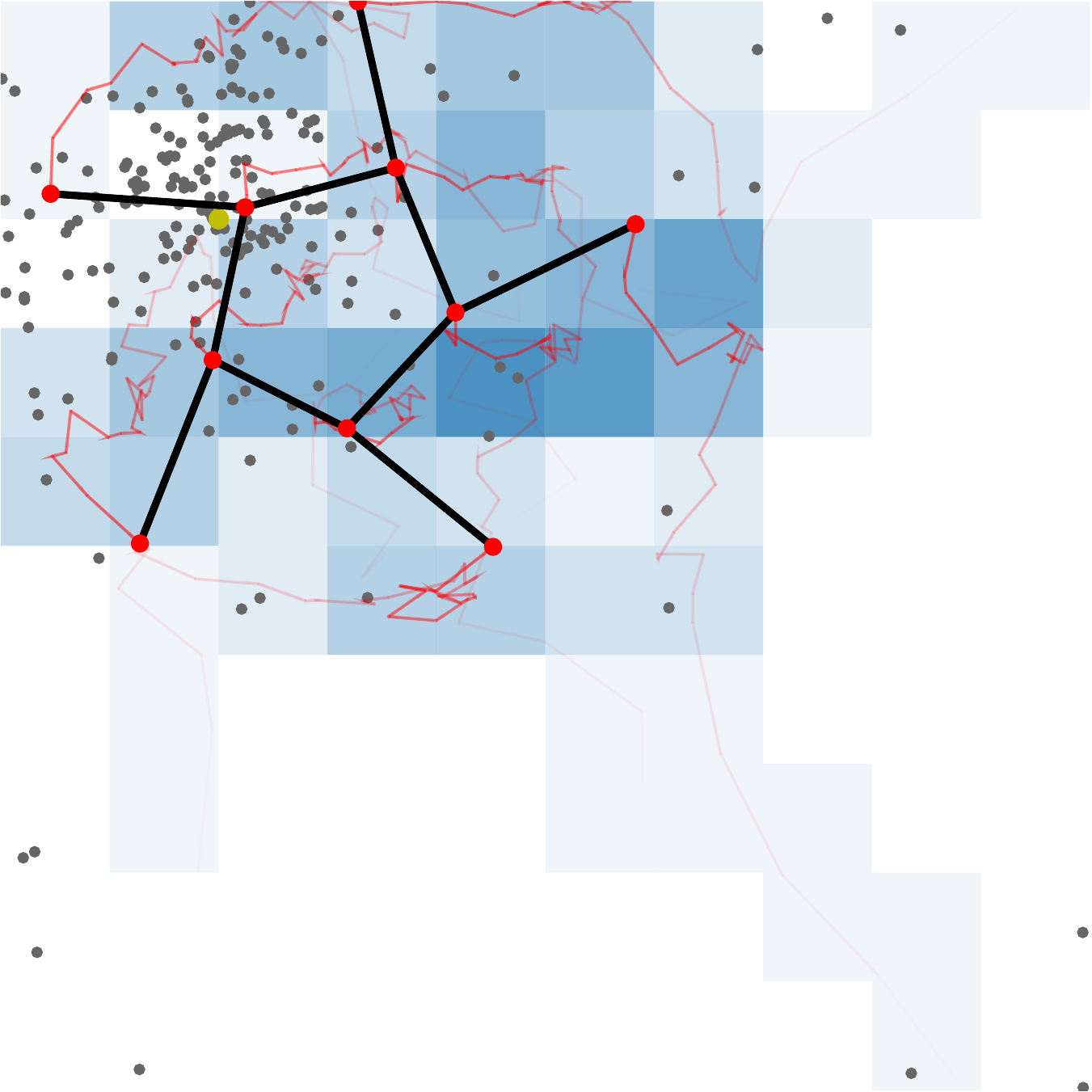}
  \href{https://www.youtube.com/watch?v=pULekr4VtcA}{LH4}
\end{minipage}
\begin{minipage}{37.2mm}
  \centering
\includegraphics[width=37.2mm]{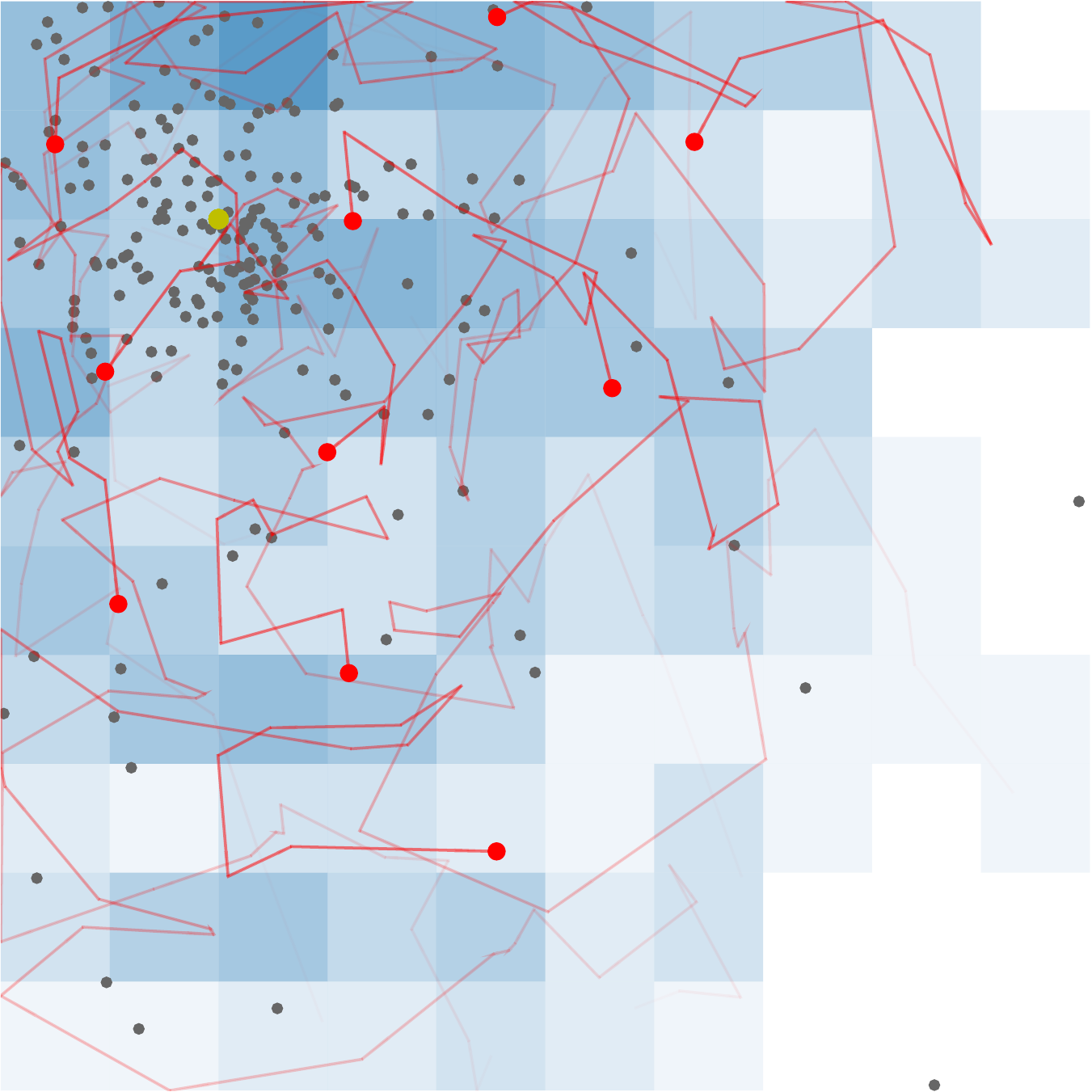}
  \href{https://www.youtube.com/watch?v=mMWvQ95hGdI}{HL4}
\end{minipage}
\begin{minipage}{37.2mm}
  \centering
\includegraphics[width=37.2mm]{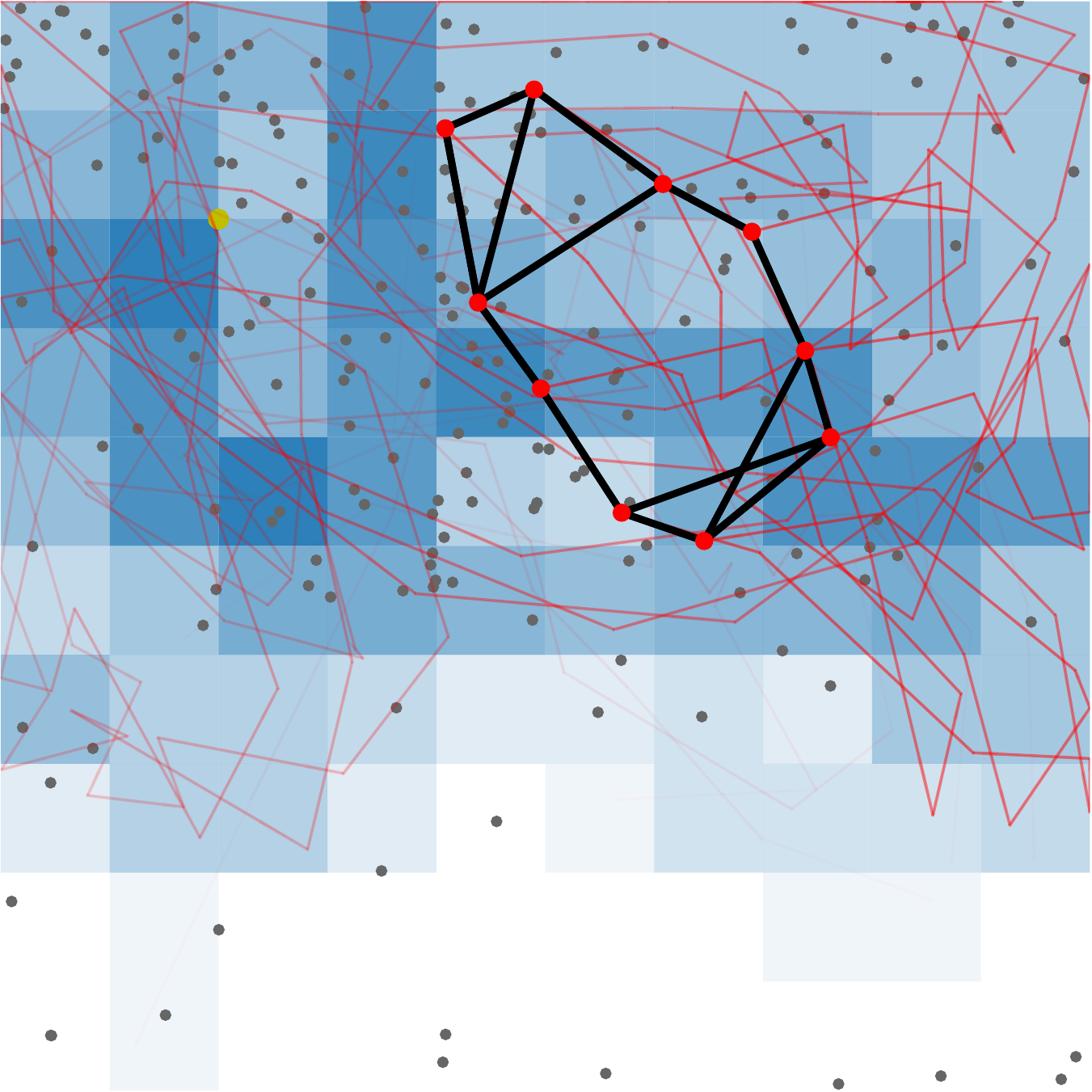}
  \href{https://www.youtube.com/watch?v=fjvY_3tgZTU}{HM4}
\end{minipage}
\begin{minipage}{37.2mm}
  \centering
\includegraphics[width=37.2mm]{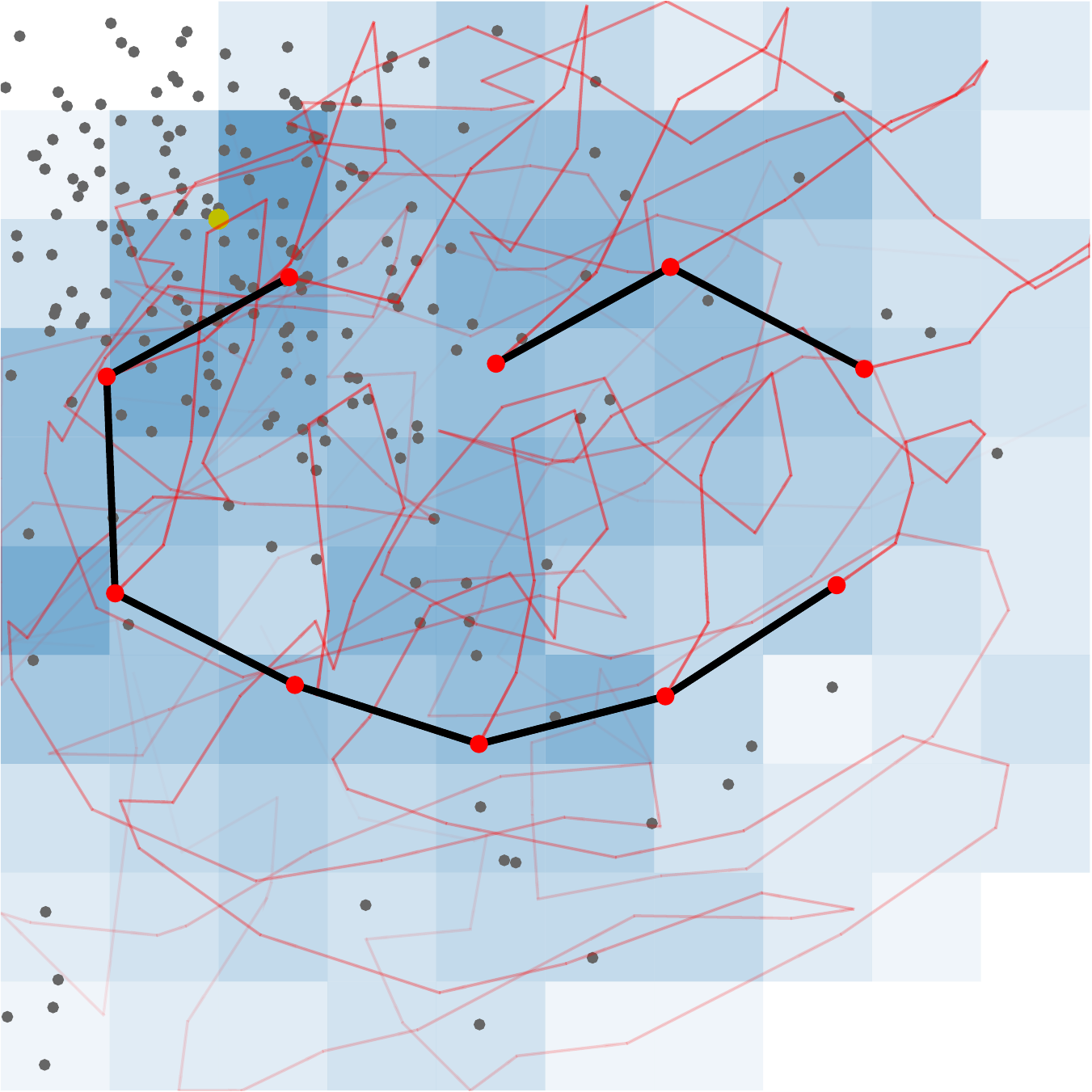}
  \href{https://www.youtube.com/watch?v=-aBeIdleerg}{HH4}
\end{minipage}
\begin{minipage}{37.2mm}
  \centering
\includegraphics[width=37.2mm]{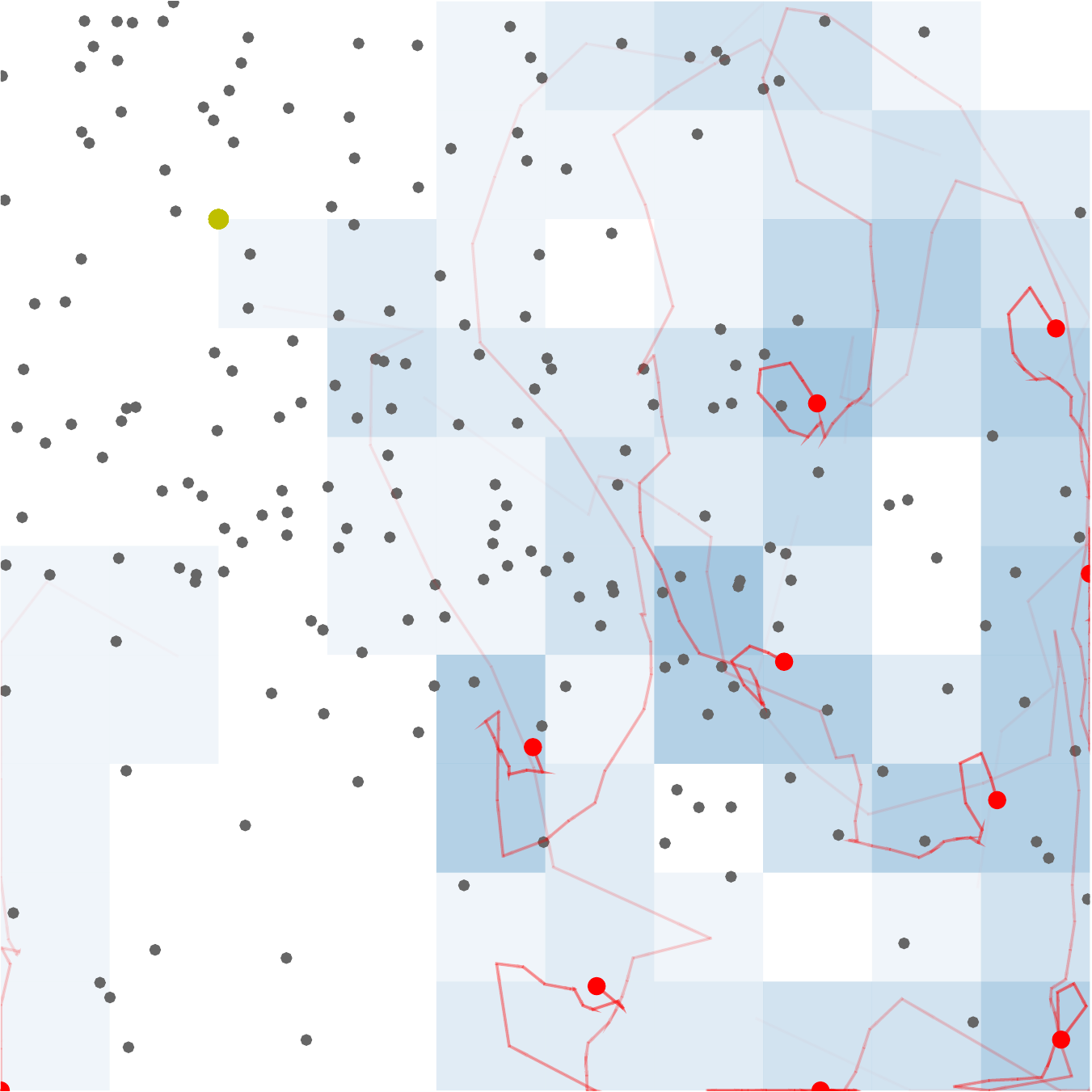}
  \href{https://www.youtube.com/watch?v=j9GHw-2QRpU}{LL7}
\end{minipage}
\begin{minipage}{37.2mm}
  \centering
\includegraphics[width=37.2mm]{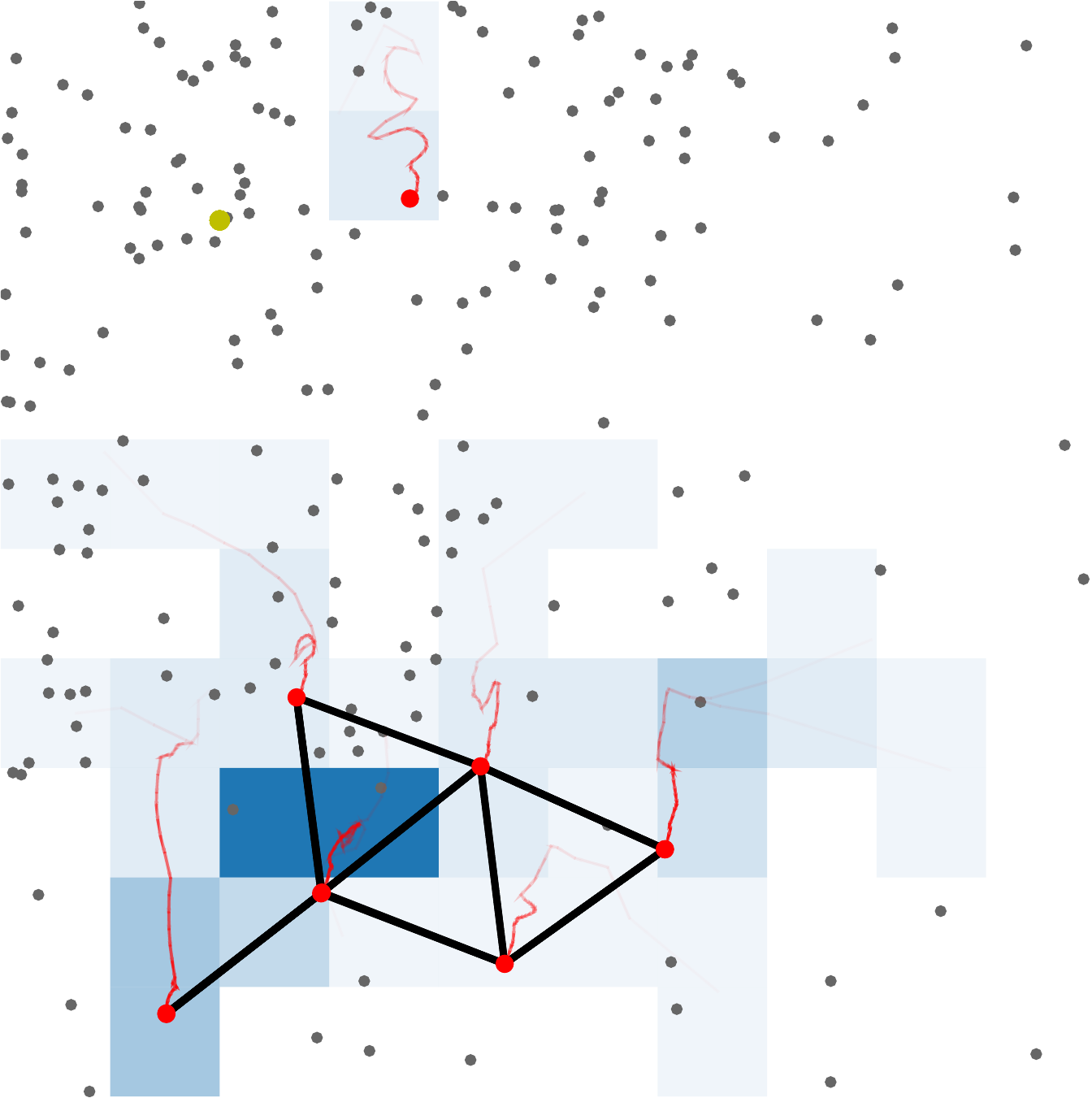}
  \href{https://www.youtube.com/watch?v=39_TvkEac-Q}{LM7}
\end{minipage}
\begin{minipage}{37.2mm}
  \centering
\includegraphics[width=37.2mm]{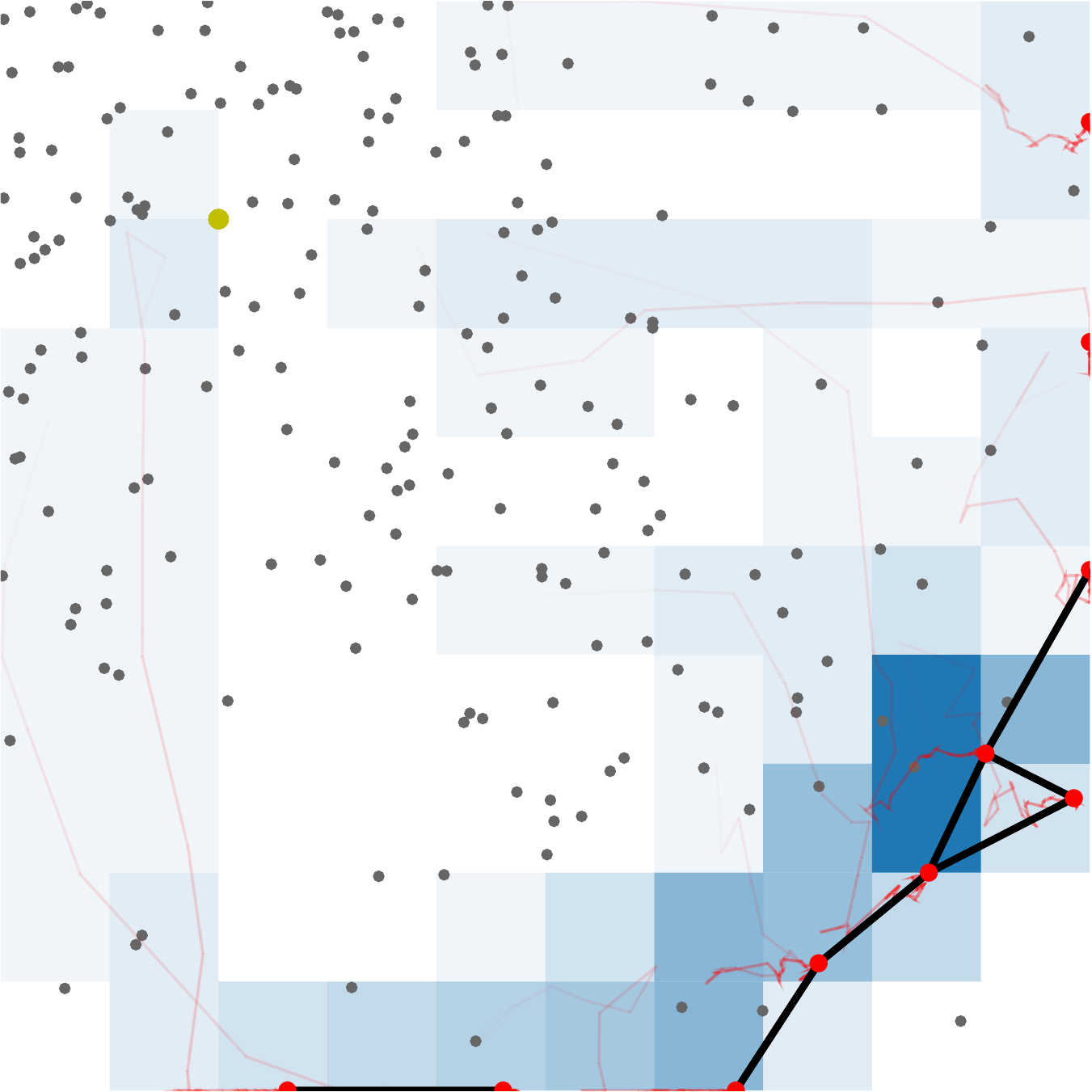}
  \href{https://www.youtube.com/watch?v=OpVn1kW6fj8}{LH7}
\end{minipage}
\begin{minipage}{37.2mm}
  \centering
\includegraphics[width=37.2mm]{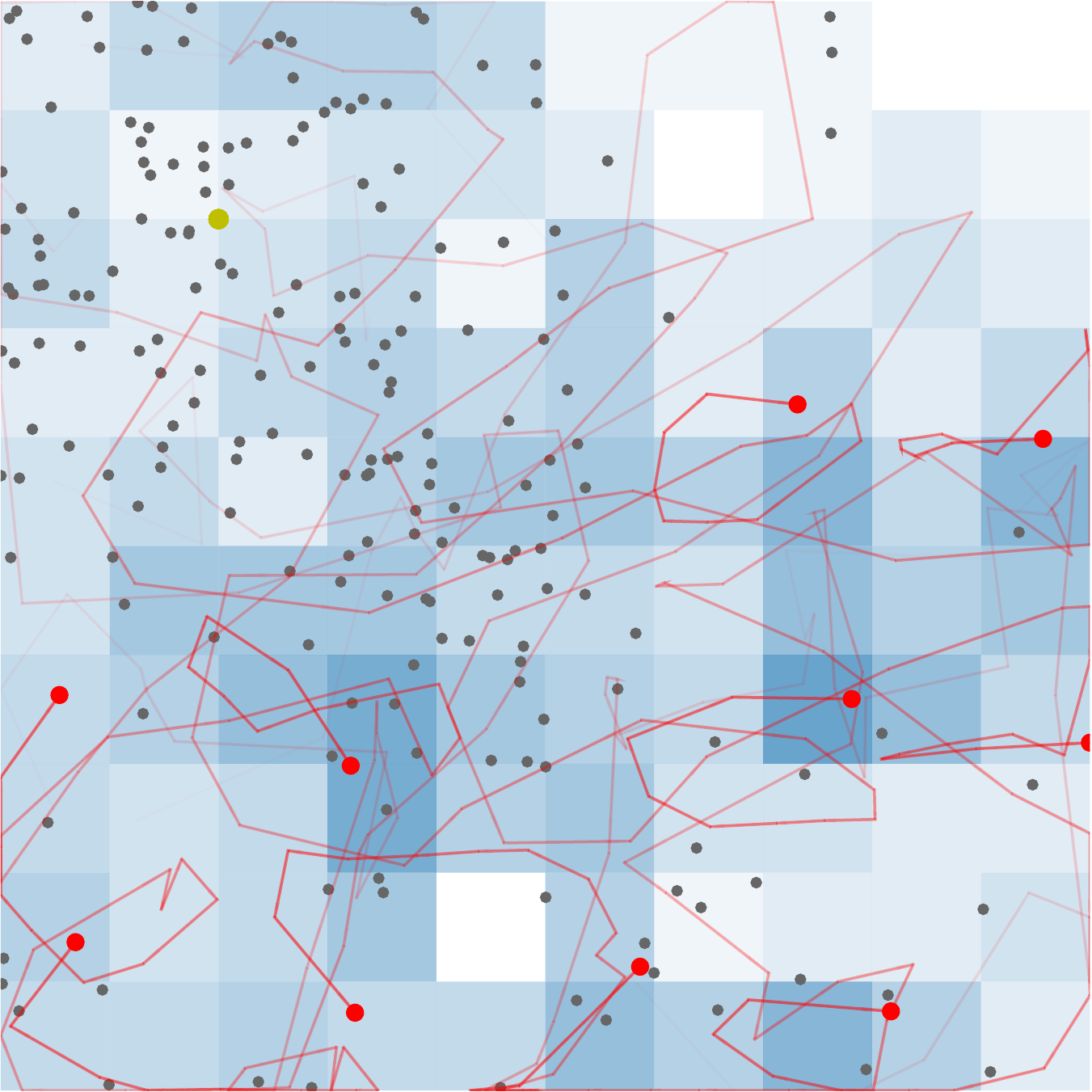}
  \href{https://www.youtube.com/watch?v=f0W3HdoiNGI}{HL7}
\end{minipage}
\begin{minipage}{37.2mm}
  \centering
\includegraphics[width=37.2mm]{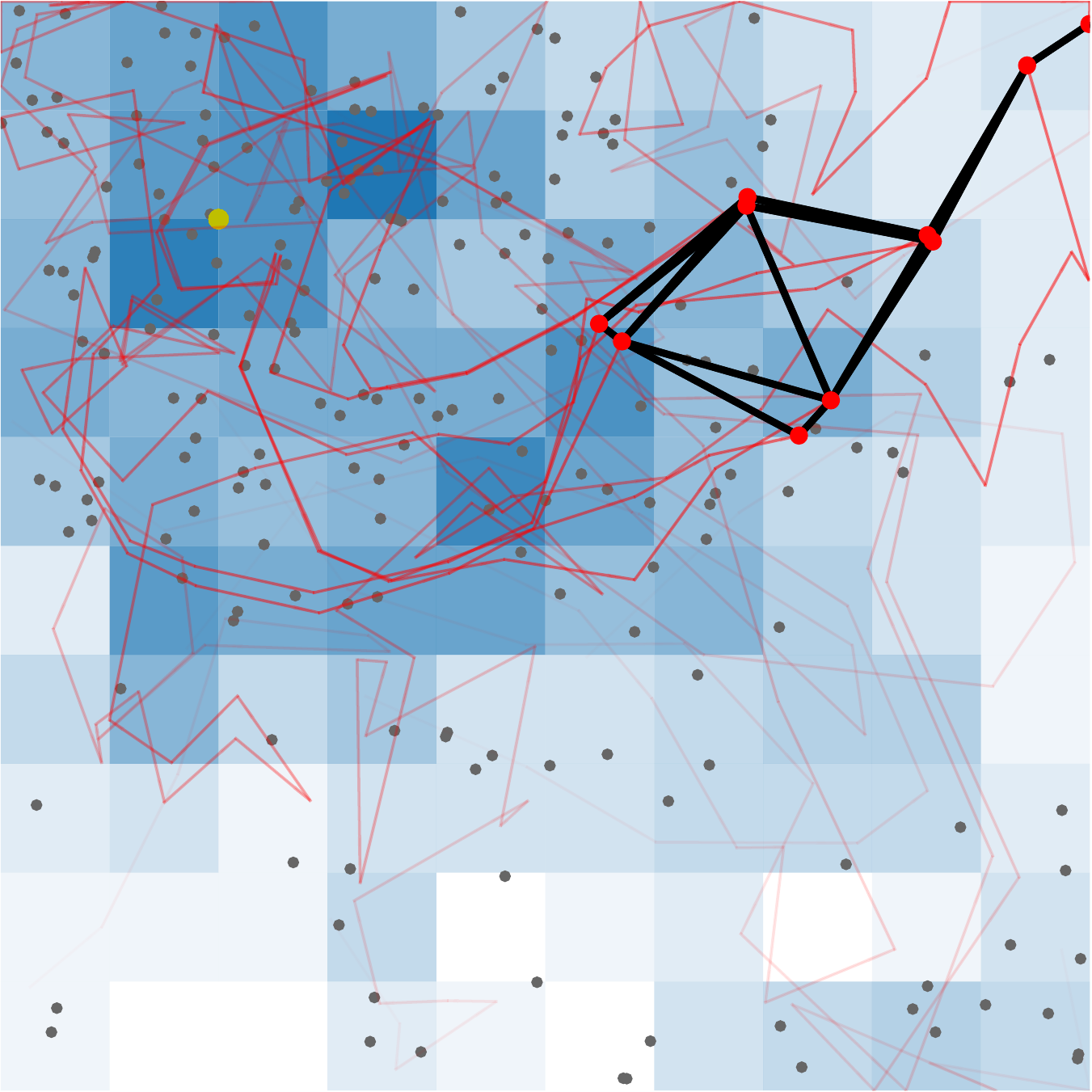}
  \href{https://www.youtube.com/watch?v=U83STfusXMY}{HM7}
\end{minipage}
\begin{minipage}{37.2mm}
  \centering
\includegraphics[width=37.2mm]{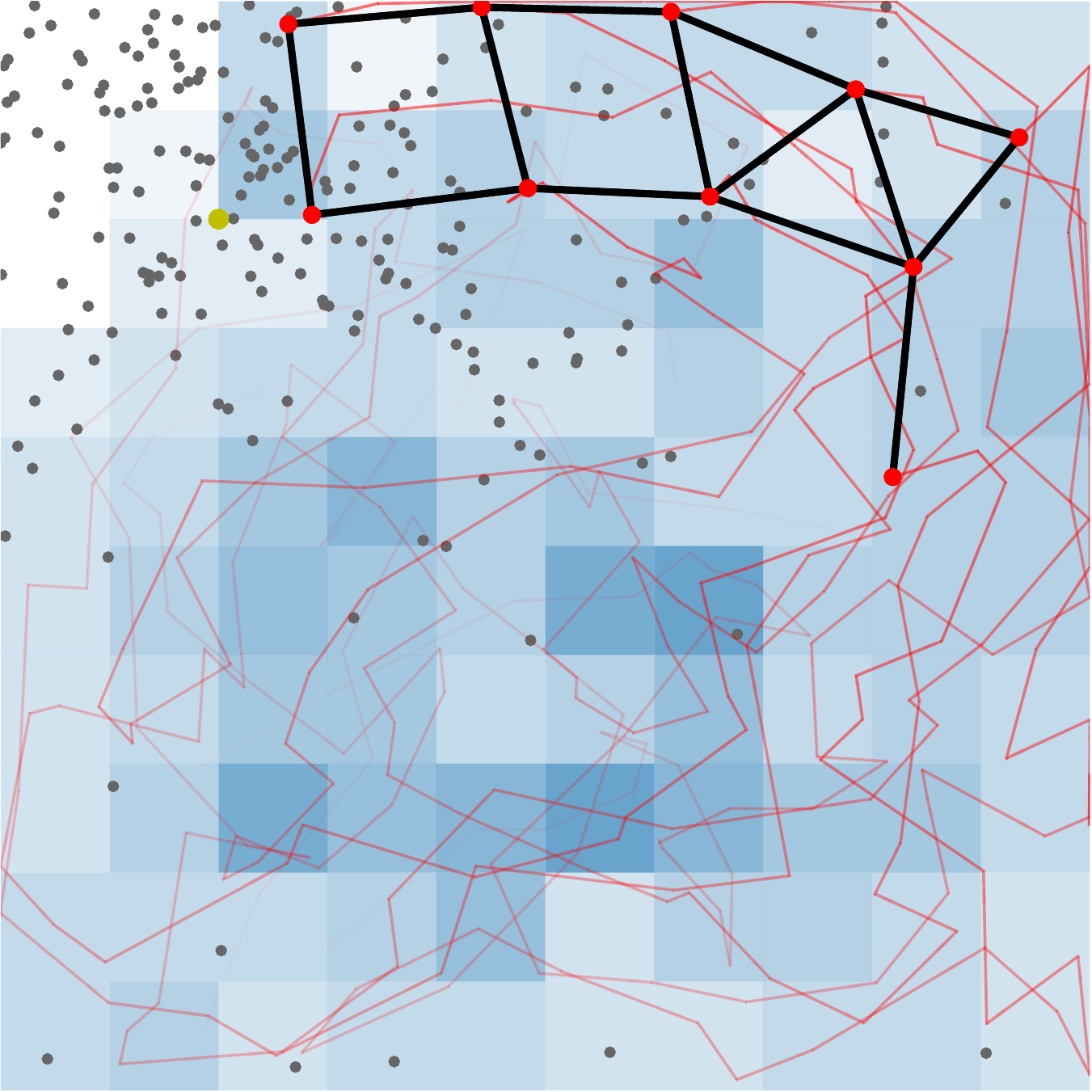}
  \href{https://www.youtube.com/watch?v=4SfK1au--jw}{HH7}
\end{minipage}

\caption{Trace plot of behaviors from Figure \ref{fig:mapelites_combined}. Labels refer to locations in repertoire. LHL for instance is a behavior with low exploration, high network coverage and low variance in geolocation predictions. Red lines indicate the path of the \glspl{uav} and black lines are connected \glspl{uav}, as determined by communication radius. Grey dots indicate location predictions. Deeper blue squares are more frequently explored. Behavior labels link to videos, and a complete overview can be found at \url{https://www.youtube.com/playlist?list=PL18bqX3rX5tT7p94T2_j2B3C4HkiMVZvY}}
\label{fig:traces}
\end{figure}

In this subsection, a subset of the evolved controllers is examined and their viability as behavior primitives are investigated. The best controllers found across all runs are stored in a repertoire (Figure \ref{fig:mapelites_combined}). From this repertoire, 16 controllers are selected by visual inspection and examined in greater detail. Selecting behaviors by visual inspection is possible because any repertoire can be flattened by slicing it  (\cite{cully2015robots}). Controllers are selected on the boundary of the feasible controller region, as this is assumed to provide the most extreme set of varied behaviors. Figure \ref{fig:mapelites_combined} shows the location of each of the solutions and Figure \ref{fig:traces} shows trace plots of the behaviors.

\begin{figure}[h]
\centering
\begin{minipage}{85.2mm}
  \centering
\includegraphics[width=85.2mm]{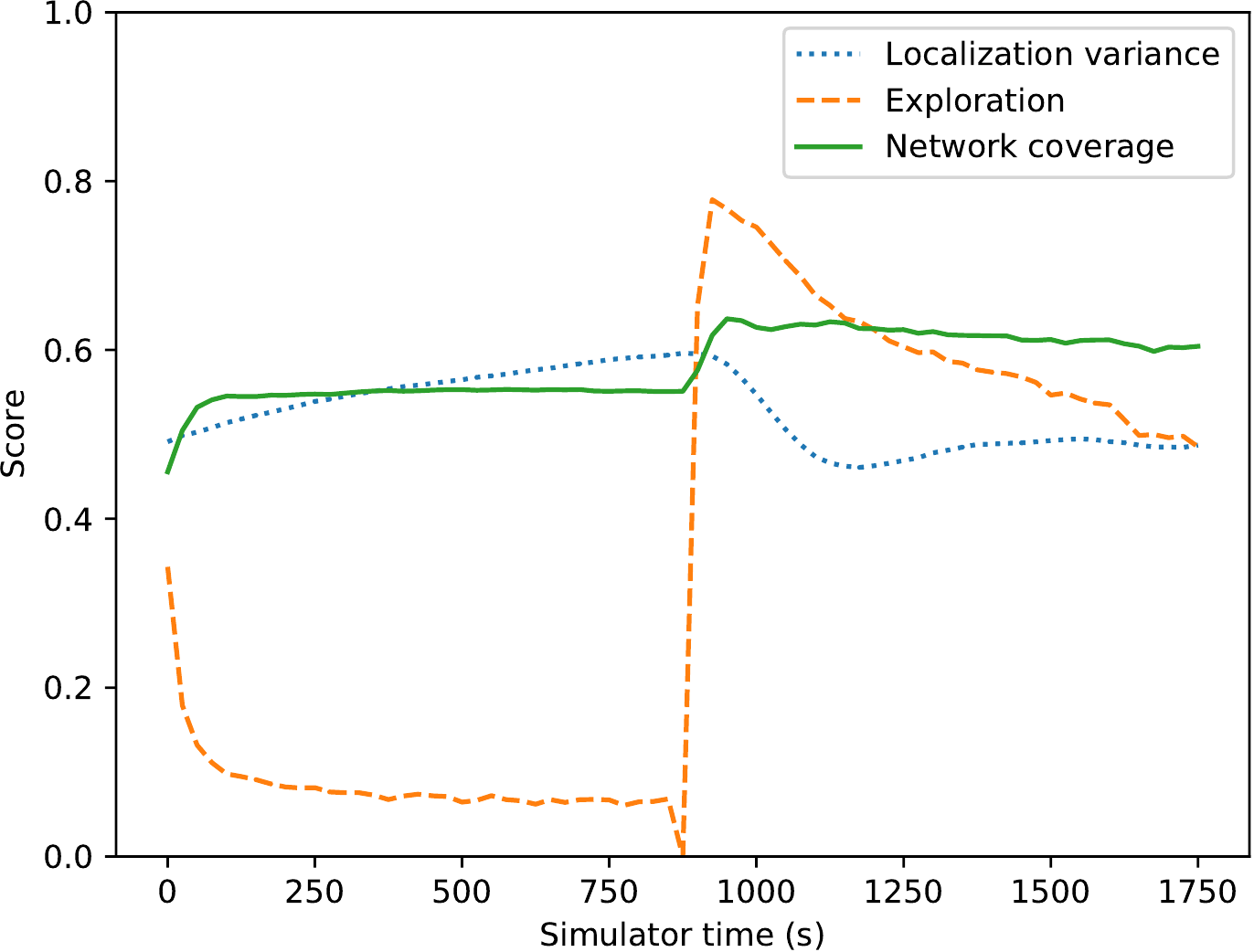}
  HM4-LM7
\end{minipage}
\begin{minipage}{85.2mm}
  \centering
\includegraphics[width=85.2mm]{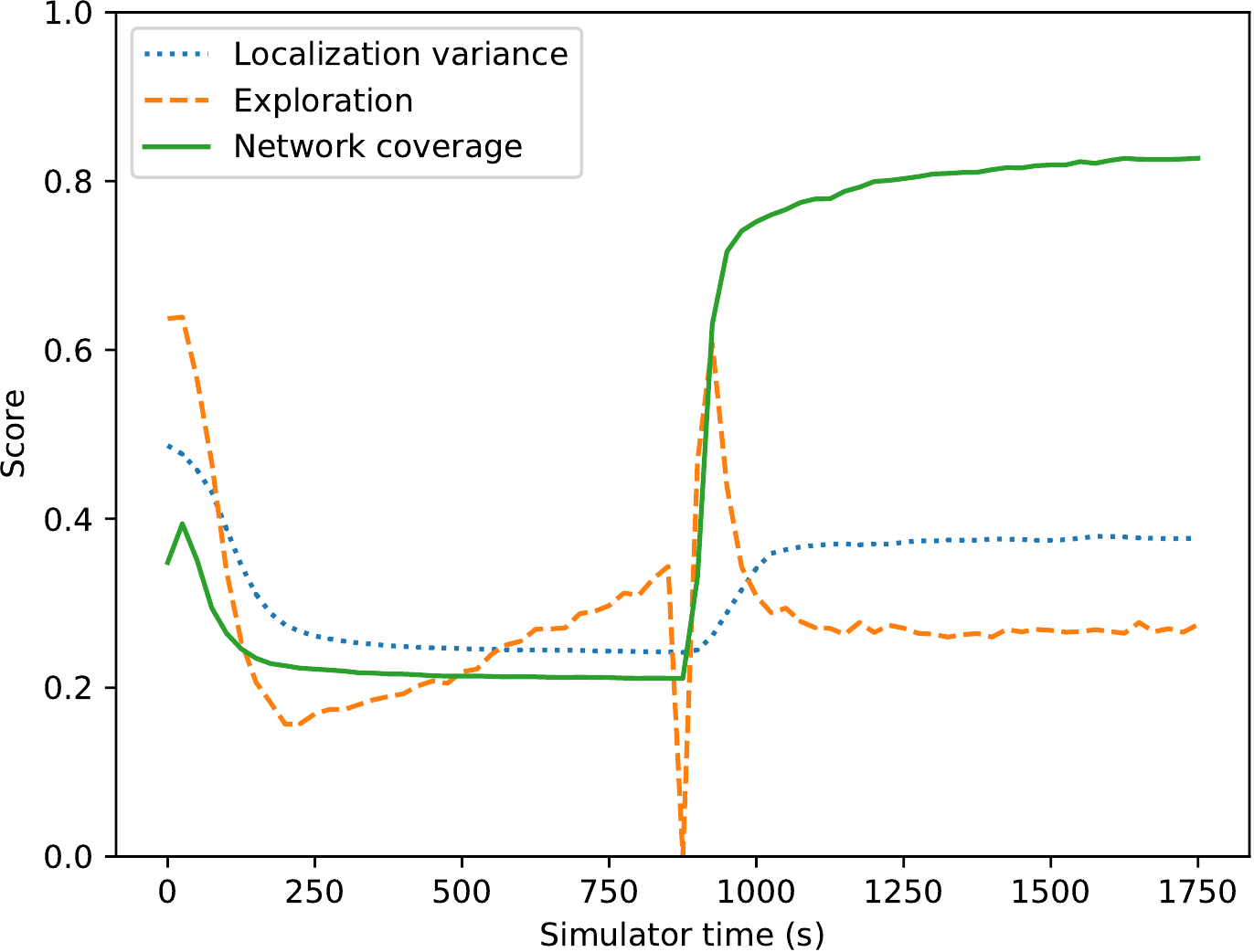}
  LH4-LLL
\end{minipage}

\begin{minipage}{85.2mm}
  \centering
\includegraphics[width=85.2mm]{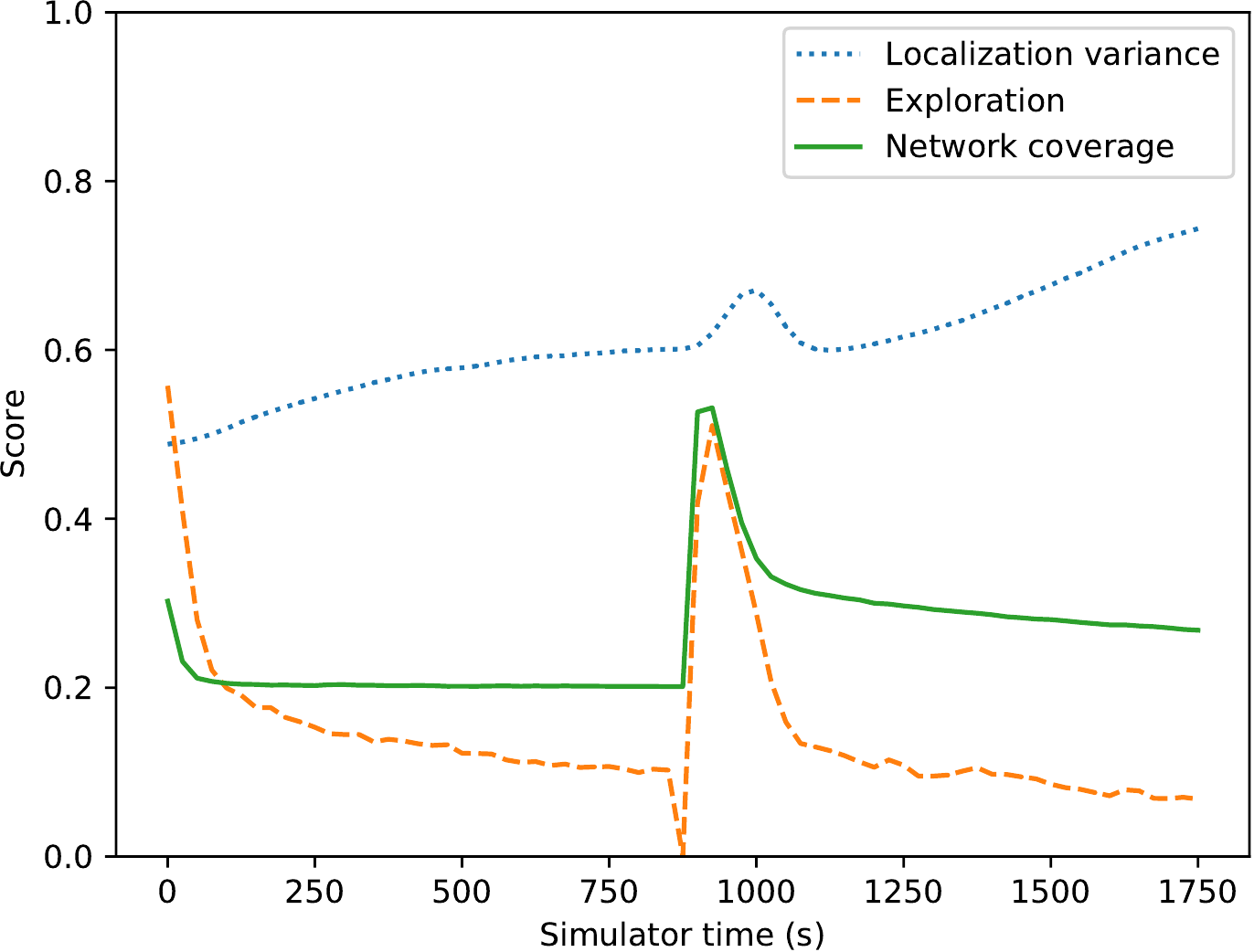}
  LHH-LL7
\end{minipage}
\begin{minipage}{85.2mm}
  \centering
\includegraphics[width=85.2mm]{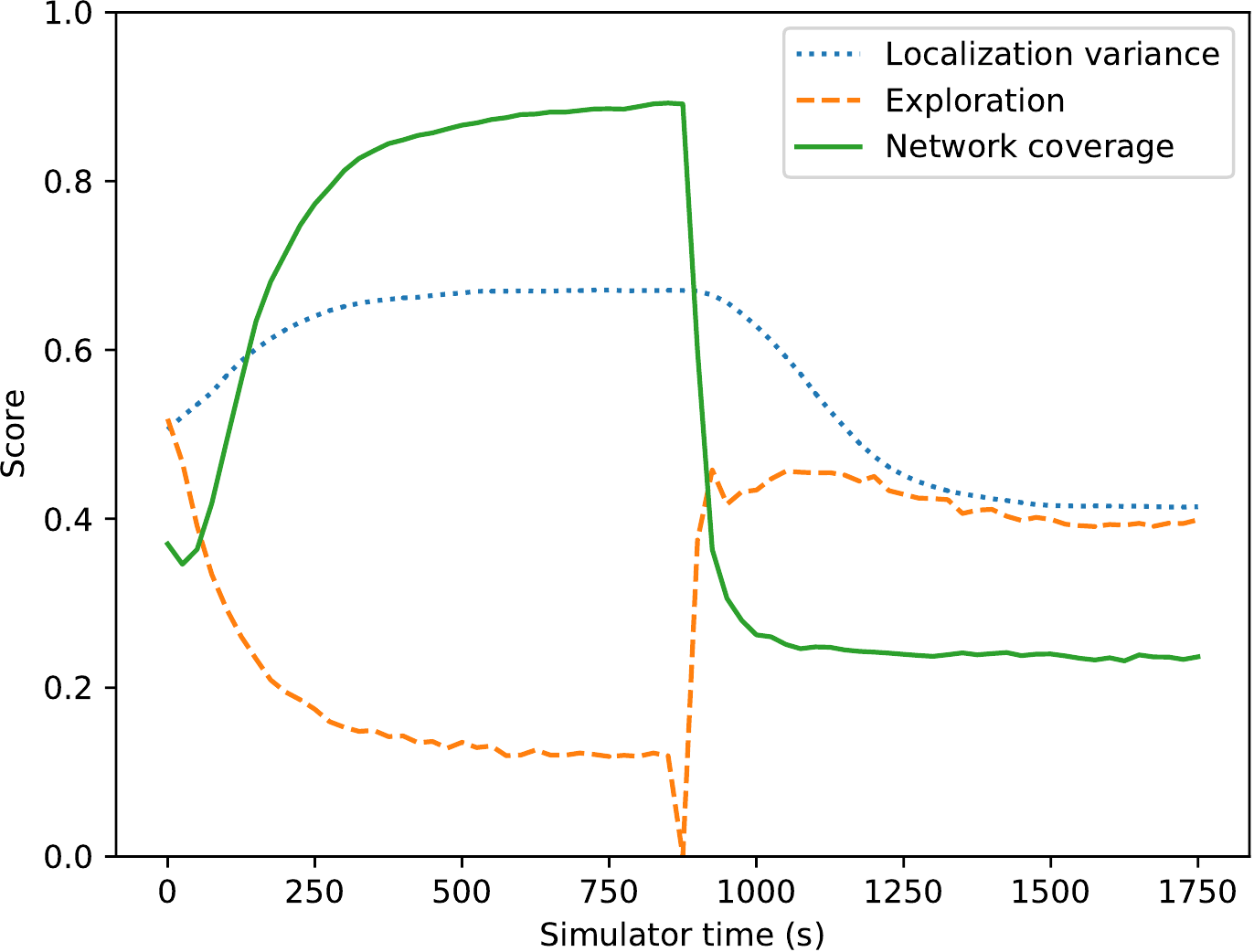}
  HL4-LH7
\end{minipage}

\caption{Four example transitions between two behaviors. Transitions happens at time 900. Graphs show averages over 1000 tests. There is a temporary reduction in exploration score during the transition due to the way exploration is measured and the agents re-organizing to the new behavior.}
\label{fig:transition}
\end{figure}

Transitioning between different behaviors shows whether the behaviors may be used as behavior primitives. The transitions between the selected controllers are examined using a surrogate metric for the exploration behavior characteristic. The exploration metric used when evolving behaviors calculates the median visitation count across all cells in the simulation area. This does not work on a short timescale as the area is too large and the median visitation count rarely gets above 0. Instead, the value of this metric is approximated using the derivative of the total visitation count calculated over the time interval since the metric was last evaluated. This gives a rough indication of how good the behaviors are in the different applications, but is subject to noise. To alleviate the noise in this measurement each transition is tested 1000 times. The average time series are shown in Figure \ref{fig:transition}. The figure shows four examples of transitions between behaviors. As far as the controller is concerned, there is no discernible difference between starting from where another behavior left off versus from a clean simulation. This is essential for these behaviors to be applicable as a set of behavior primitives and is what could enable the operator to change the behavior of the swarm on the fly.

\subsection{Investigating the effect of noise in MAP-elites}


MAP-elites is a greedy algorithm. Every time a solution is mutated it is kept as a part of the repertoire if it fills a void where there previously was no solution, or it is better than the existing solution. It is an excellent property to maintain diversity and allows for better exploration of the search landscape, but also poses a challenge. Many common metrics or fitness functions used in evolutionary optimization are stochastic. This applies to both fitness metrics and behavior characteristics. If a stochastic variable has high variance, but a low mean it still might outcompete a stochastic variable with high mean and low variance. In the case of these experiments, a behavior might get a lucky draw from the metrics used to evaluate performance, resulting in an inferior solution being chosen over a superior solution. This is a challenge, as in many cases, the lesser variability and higher mean may be preferable to the lower mean and greater variability.

\begin{figure}[h]
\centering
\begin{minipage}{58mm}
  \centering
\includegraphics[width=58mm]{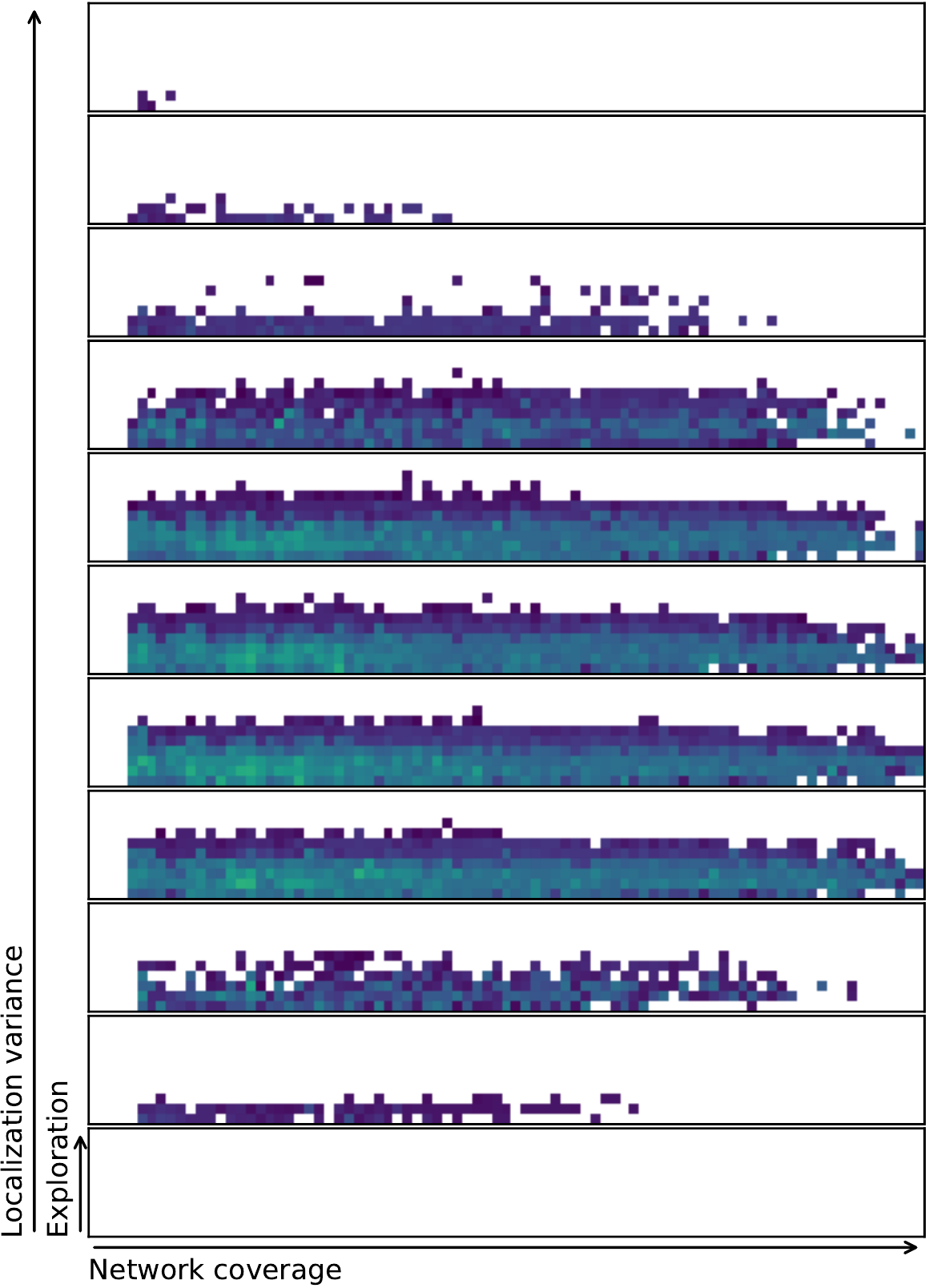}
  1-evaluation (2937)
\end{minipage}%
\begin{minipage}{58mm}
  \centering
\includegraphics[width=58mm]{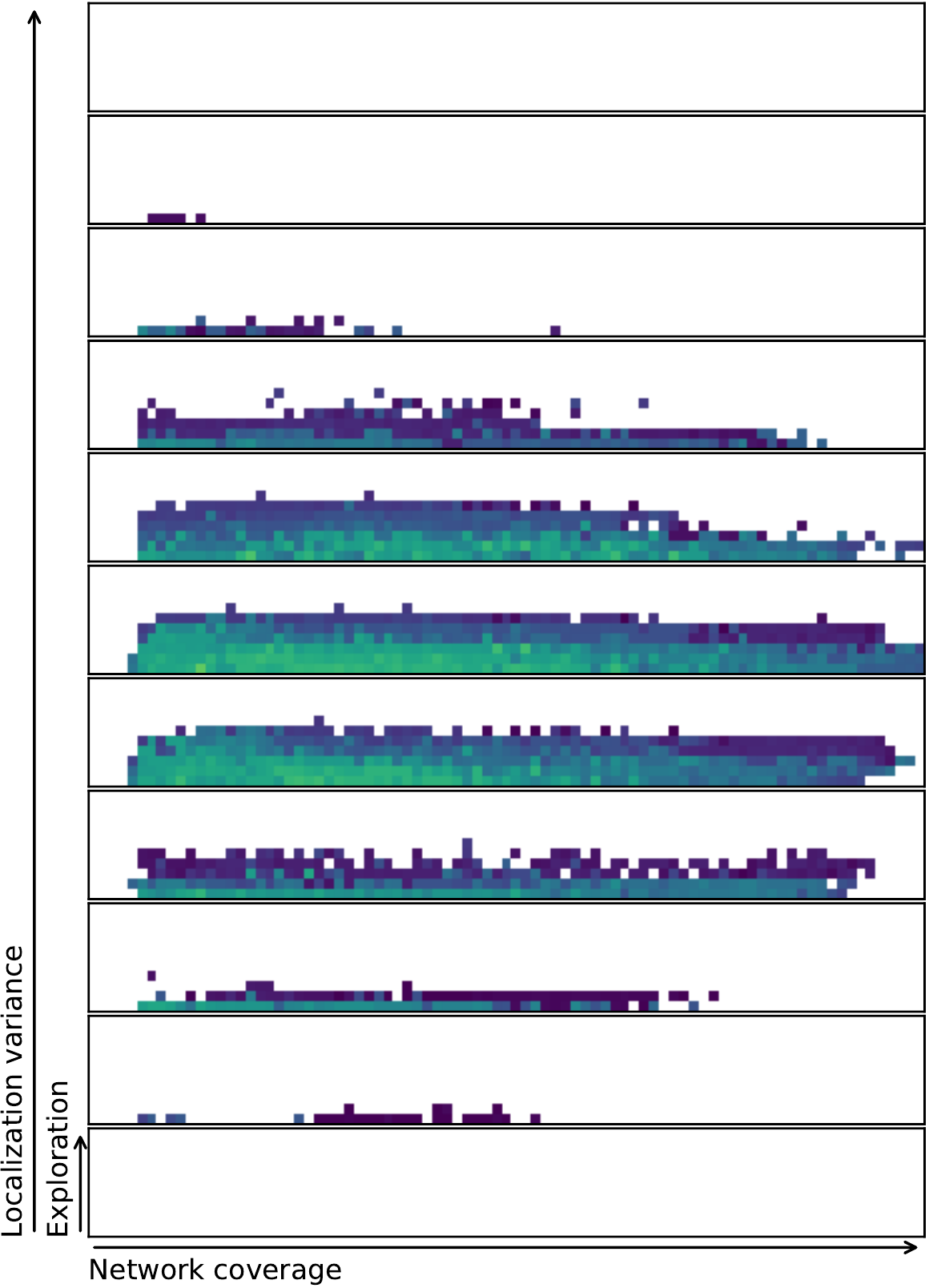}
  5-evaluations (1957)
\end{minipage}%
\begin{minipage}{58mm}
  \centering
\includegraphics[width=58mm]{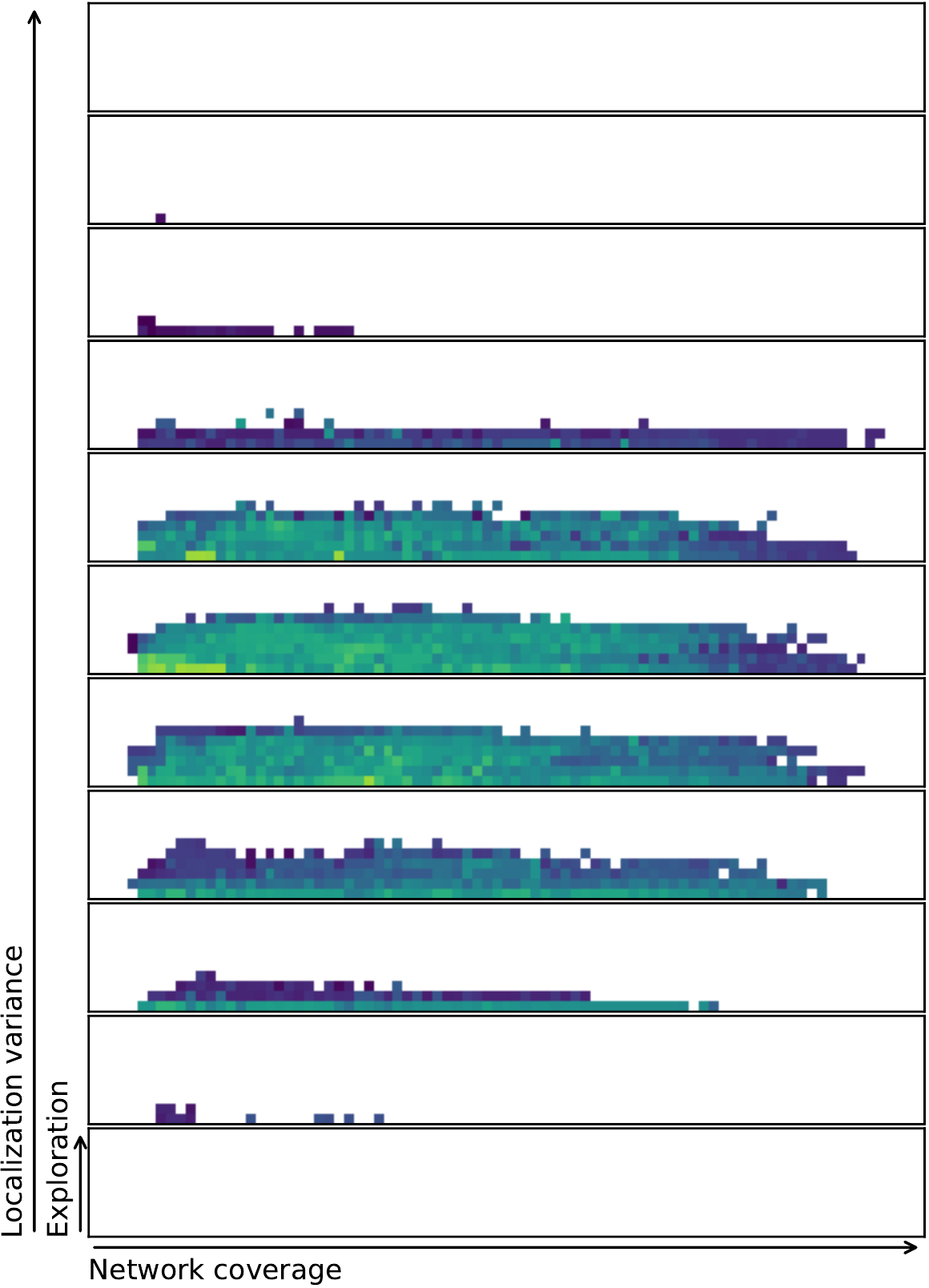}
  10-evaluations (1841)
\end{minipage}%
\vspace{1em}
\begin{minipage}{58mm}
  \centering
\includegraphics[width=58mm]{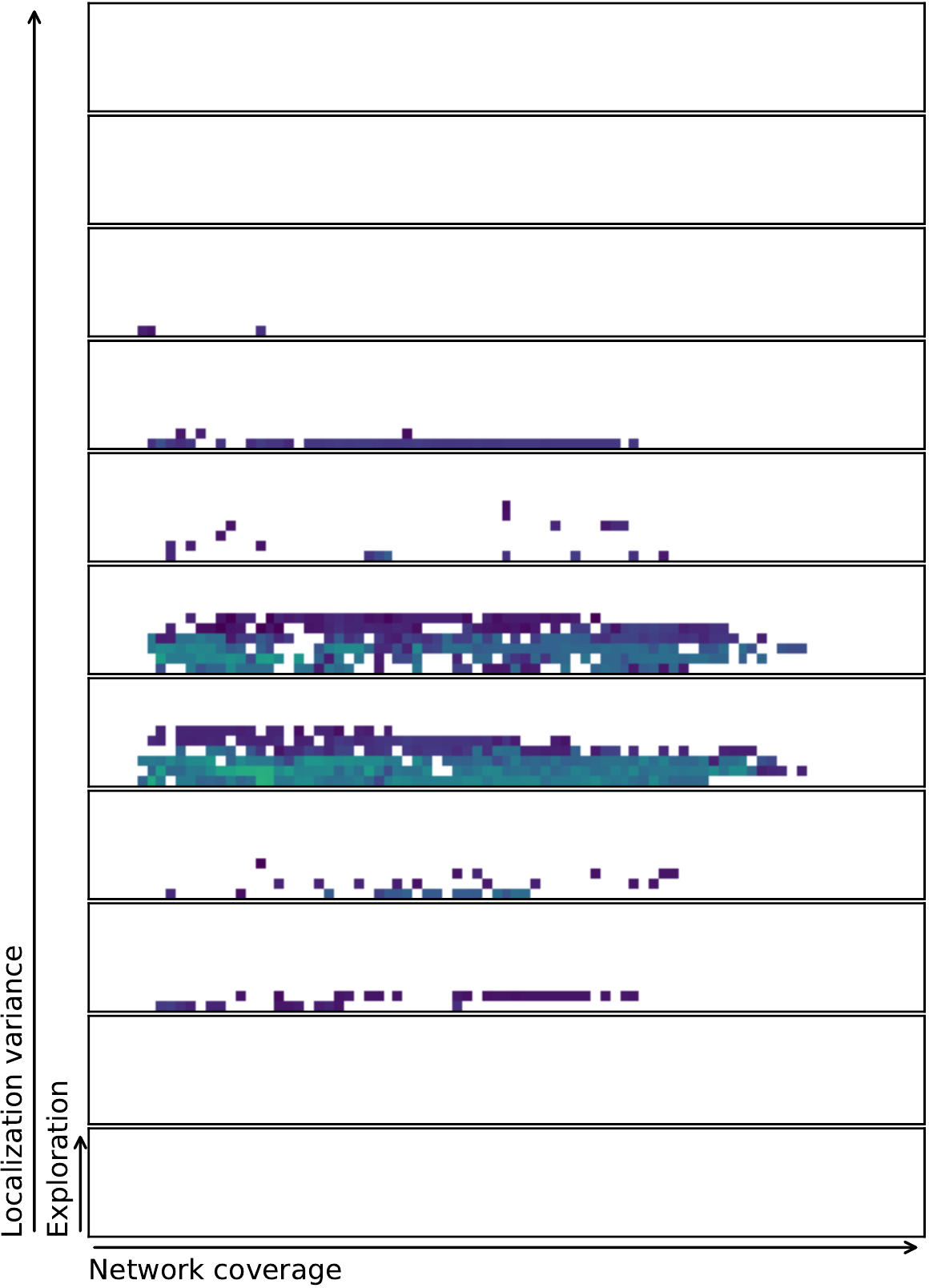}
  Re-eval 1-eval. (700)
\end{minipage}
\begin{minipage}{58mm}
  \centering
\includegraphics[width=58mm]{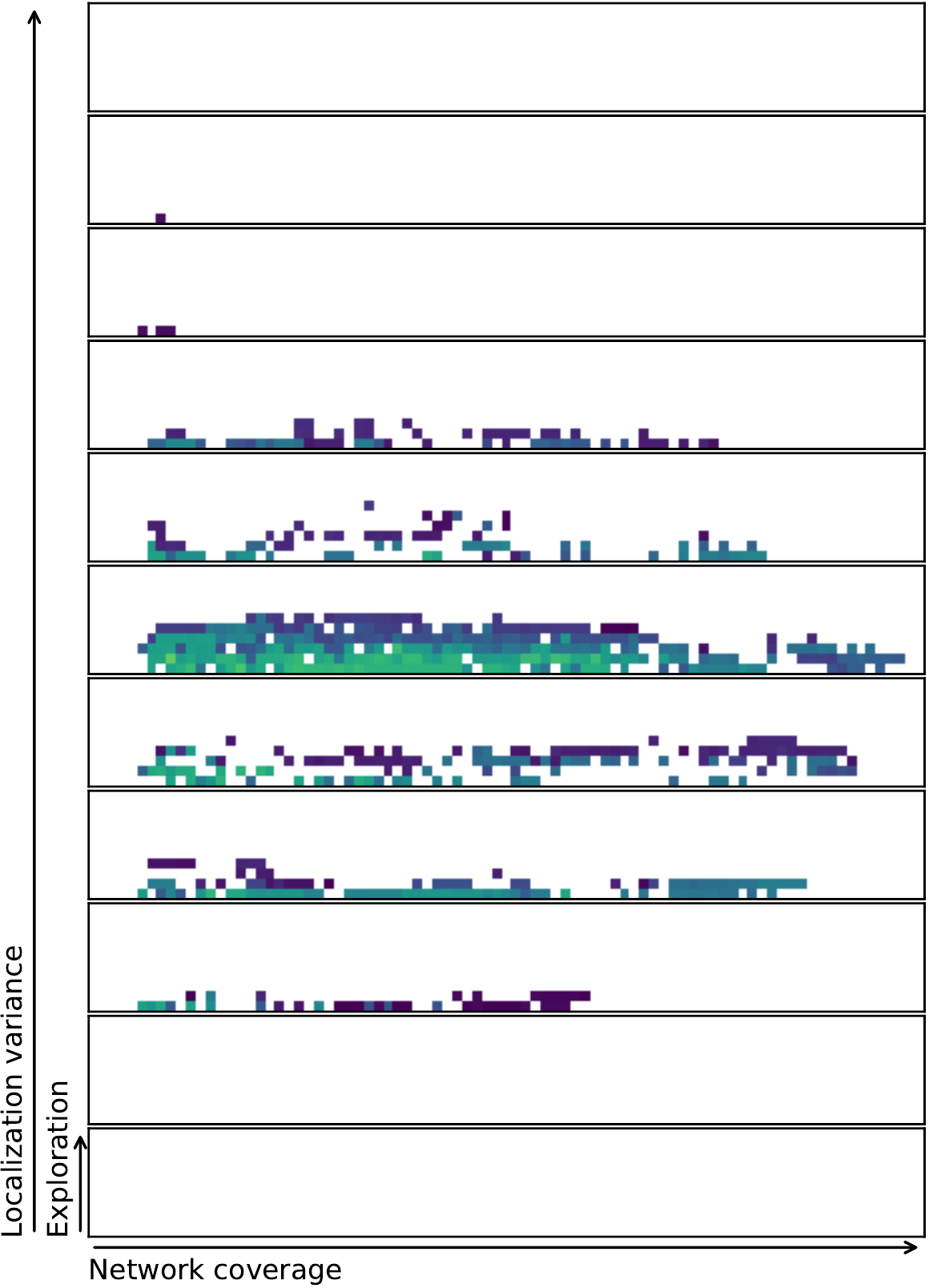}
  Re-eval 5-eval. (724)
\end{minipage}
\begin{minipage}{58mm}
  \centering
\includegraphics[width=58mm]{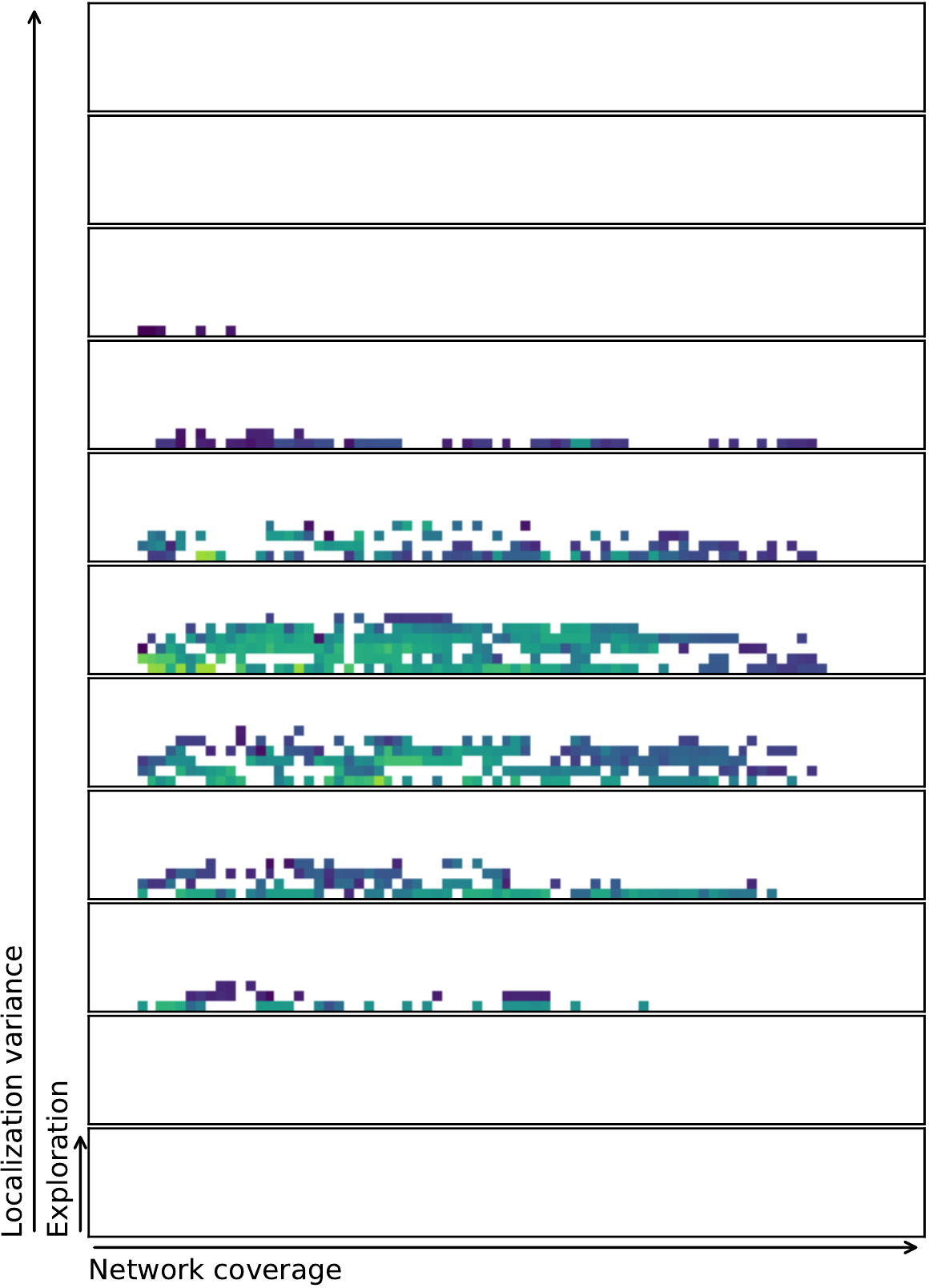}
  Re-eval 10-eval. (743)
\end{minipage}
\caption{Original (top) and re-evaluated repertoires (bot) for runs using an average of 1, 5, and 10 evaluations per controller in the initial repertoire evolution. Re-evaluation with 100 evaluations per candidate solution results in a large reduction in repertoire size. The number of solutions in each repertoire is shown in parentheses.}
\label{fig:mapelites_noise}
\end{figure}

In order to test if a repertoire is reproduceable an entire repertoire is re-evaluated using more evaluations per solution or behavior. This gives a clear visual indication of whether the solution stays in the same characteristics bin or moves around. Fig \ref{fig:mapelites_noise} and Table \ref{table:mapelites_noise} shows that there is a noticeable reduction in the number of solutions in the repertoires as they are re-evaluated. The figure shows an overview of repertoires evolved and re-evaluated with a single, 5 evaluations and 10 evaluations per controller. It is important to note that using more evaluations in the initial evolution results in a smaller repertoire. However, this smaller repertoire is likely closer to the true shape of the space of all feasible swarming behaviors. Using only a single evaluation results in a repertoire with 2937 solutions, while 5-evaluations results in only 1957 solutions. The further reduction with 10-evaluations is much smaller, with the final repertoire having 1841 solutions.

\begin{table}[h]
\caption{Overview of results of re-evaluation of controller repertoires.}
\begin{tabular}{llll}
 & 1-eval repertoire & 5-eval repertoire & 10-eval repertoire \\
Original & 2937  & 1957 & 1841 \\
Re-eval 20-eval & 823 (28.0\%) & 857 (43.8\%) & 859 (46.7\%) \\
Re-eval 50-eval & 744 (25.3\%) & 759 (38.8\%) & 800 (43.4\%) \\
Re-eval 100-eval & 700 (23.8\%) & 724 (37.0\%) & 743 (40.4\%) \\
\end{tabular}
\label{table:mapelites_noise}
\end{table}

It is important to note that re-evaluating using 20 evaluations per solution cannot be compared to evolving a repertoire with 20 evaluations per solution. During the evolutionary process, a total of 40 000 solutions are tested. During re-evaluation, only the ~2000-3000 solutions in the repertoire are re-evaluated. As such, it is natural that the re-evaluated repertoire contains a lot fewer solutions as it was not given time to search for solutions to fill all the characteristics bins. 

Increasing the number of evaluations seems to have an effect by producing a repertoire that is more correct, or closer to the true underlying shape. However, the effect is also diminishing: going from 5 to 10 evaluations has little effect. Therefore all other experiments in this work used 5-evaluations per individual.

Table \ref{table:mapelites_noise} shows that the number of evaluations used in the re-evaluation step is not that important. The greatest reduction in repertoire size is found when using the highest number of evaluations in the re-evaluation step (100 evaluations per solution). However, the effect of increasing the number of evaluations on the robustness in the initial repertoires can also be clearly seen when re-evaluating the repertoire with only 20 evaluations per individual.  

\begin{figure}[h]
\centering
\includegraphics[width=0.9\textwidth]{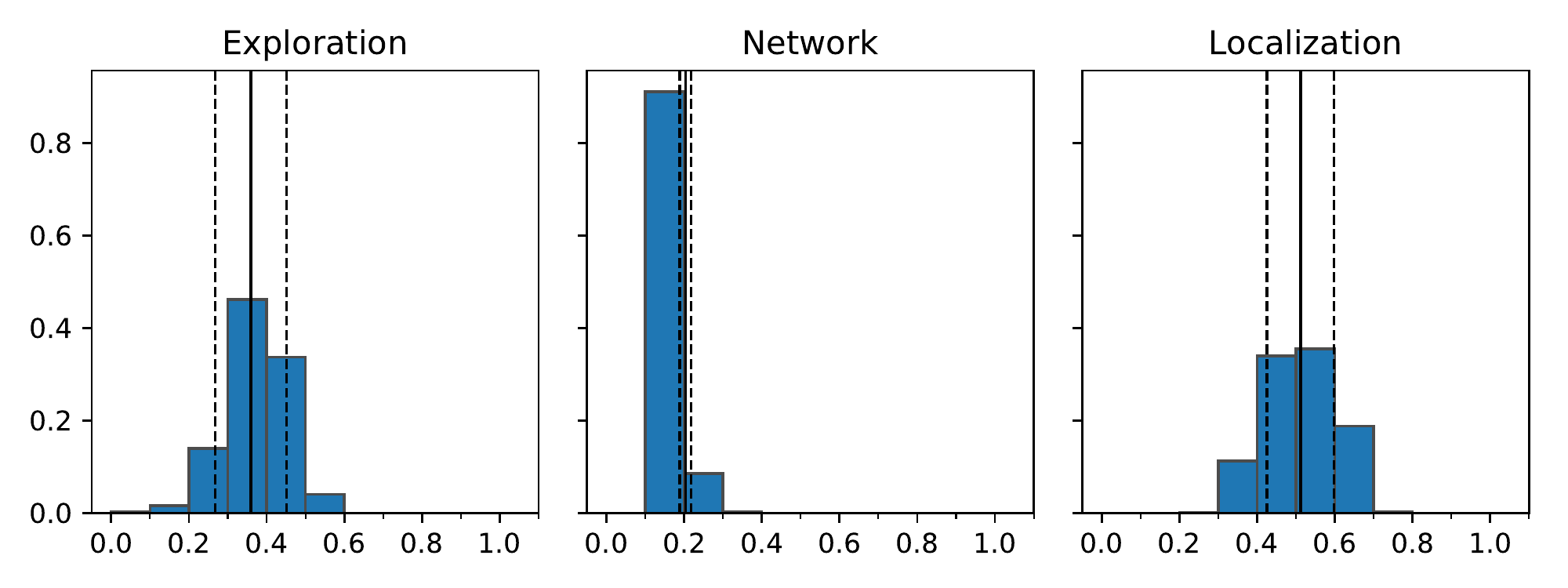}
\caption{Example of a probability distribution of characteristics over 1000 simulations of the same controller. Counts are normalized to sum to one across all 10 bins for each characteristic. The solid line is the mean of samples, dotted lines indicate one standard deviation.}
\label{fig:fitness_variation}
\end{figure}

To further investigate the challenge of reproducing behaviors and repertoires a single behavior (HL7) is re-evaluated 1000 times and the probability distribution of the behavior characteristics are shown in Figure \ref{fig:fitness_variation}. In the experiments conducted in this paper the fitness can be computed from the genome deterministically, so only the three behavior characteristics (exploration, network coverage and localization variance) are reviewed.

Figure \ref{fig:fitness_variation} shows that the distributions for each of the three metrics are fairly well behaved and resembles a normal distribution in most cases. Variation, as indicated in the figure, can cause a behavior to seemingly move between characteristics bins when re-evaluated, this is believed to the primary cause of the reduction in size of the repertoire.

\begin{figure}[!htb]
\centering
\includegraphics[width=0.7\textwidth]{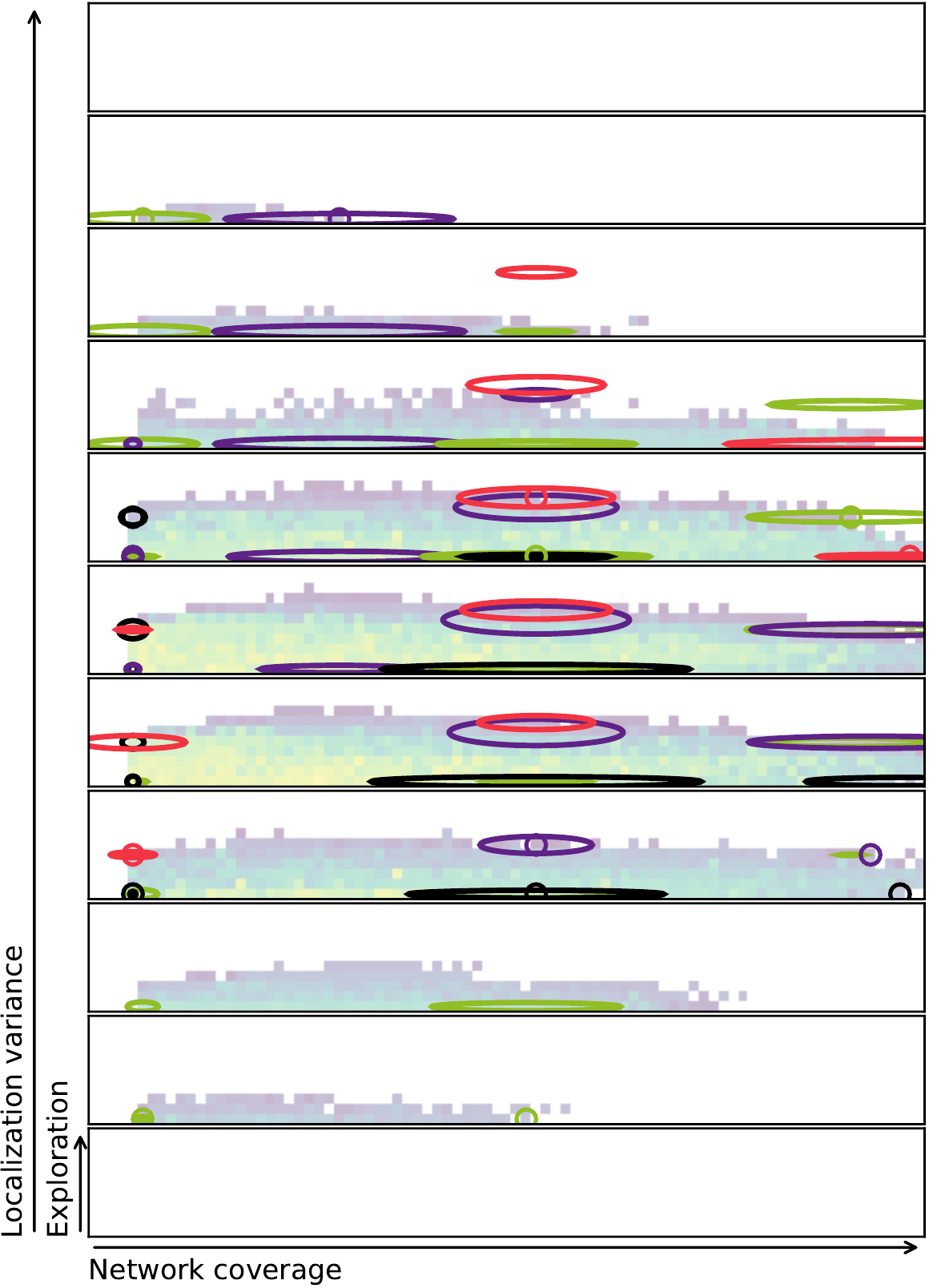}
\caption{Repertoire with uncertainty ellipses plotted for selected behaviors. The small circles indiciate the location of the behavior in the final repertoire. Slices of the uncertainty ellipse are shown in the same color as the circle indicating the behavior. Note that behaviors at the edge of the repertoire also often are at the edge of the uncertainty ellipse. This indicates that behaviors with these characteristics may be hard to come by. }
\label{fig:uncertainty}
\end{figure}

A challenge that remains is to successfully quantify the properties of the solutions. If the measure has stochastic properties, the exact same solution might fit in multiple bins in the repertoire. This again means that there is uncertainty whether the evolutionary method captures the true shape of the behavior space or not. To visualize this, Figure \ref{fig:uncertainty} shows a combined repertoire over 8 evolutionary runs with uncertainty ellipses plotted as slices in 3D. As can be seen from the figure, the behavior is not always found at the center, or mean, of the distribution. The uncertainty ellipses (one std.dev.) are estimates generated by evaluating each of the selected controllers 1000 times and calculating variance and mean over these runs.

Figure \ref{fig:uncertainty} shows how solutions might jump between characteristics bins if re-evaluated. It also indicates that many of the solutions that are in the repertoire are at the very edge of the potential range of values that the characteristics may take on. This suggests the idea that MAP-elites might be biased towards accepting solutions that have a high variance, as they sometimes get lucky and provide a solution for a hard to reach characteristics bin.

\subsection{Ablation study of the effect of controller inputs}

The proposed controller has 8 inputs that determine the action of a swarm agent at any given time. In previous works (\cite{engebraaten2018evolving}), a simpler parametric controller with only 4 inputs was employed, as well as a controller with only scalar weights, which did not enable the agents to evolve holding distance type behaviors. In this work, the distance and direction to another 3 neighbors was added. This was done in the interest of improving the performance of the controller in the 3 given applications. To quantify the effect of this change and the ability of the evolutionary process to find good swarm behaviors, an ablation study is performed. Individual inputs are disabled, which allows the effect of each input to be examined separately. Ablation refers to the selective disabling or removal of certain parts or a larger object in order to investigate the effect this might have.

\begin{figure}[h]
\centering
\includegraphics[width=0.7\textwidth]{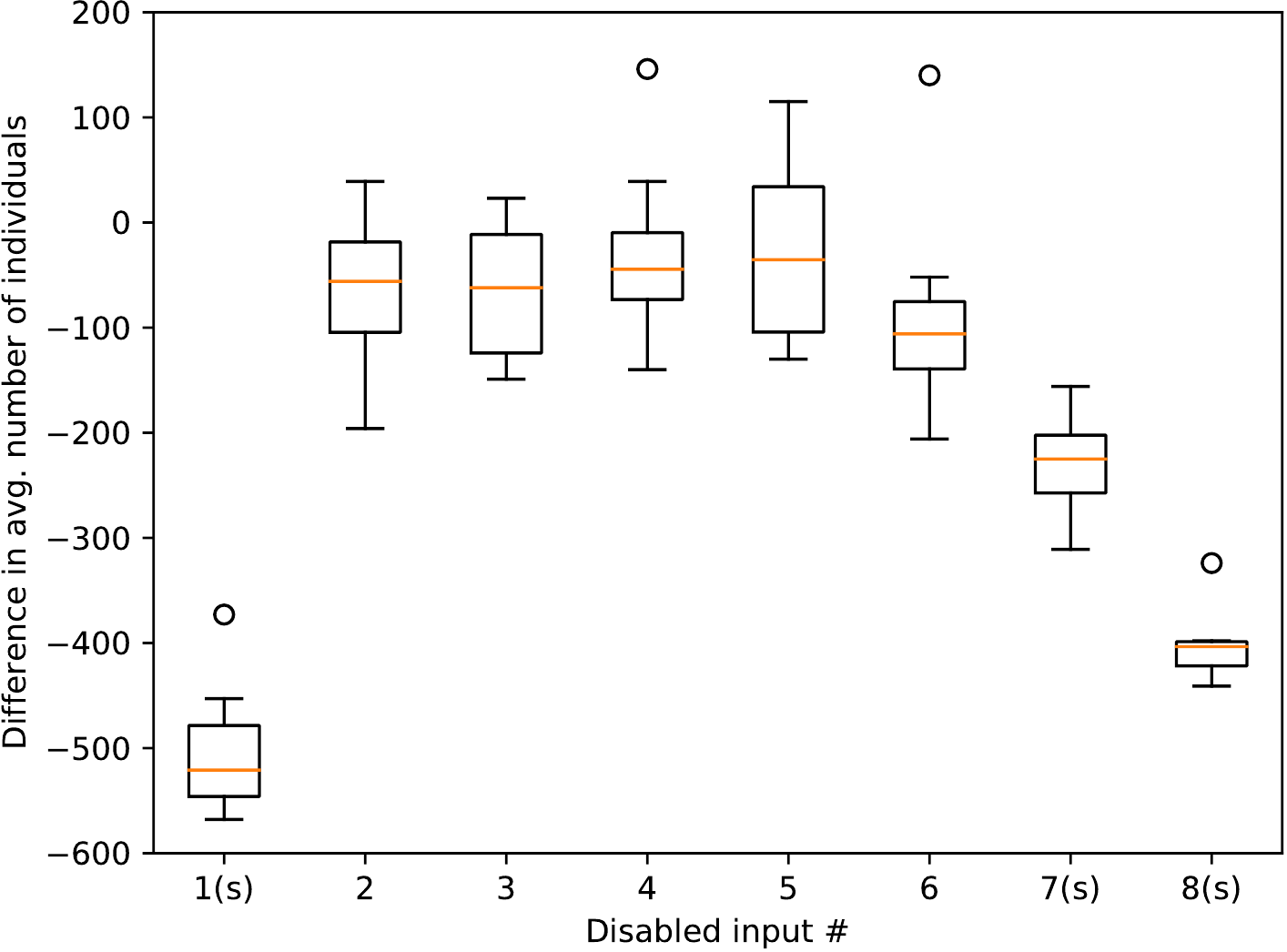}
\caption{Difference in individual count when disabling each input to the swarm controller. An (s) after the X-axis label indicates statistical significance (P \textless 0.01) in Rank-Sum test, comparing against all inputs enabled. Tests remain significant when corrected with Bonferroni for potential multiple comparison errors. Circles indicate outliers outside the range of the whiskers. This box plot uses whiskers that may extend up 1.5 times of the interquartile range; any data points outside this range are considered outliers.}
\label{fig:count_disabled}
\end{figure}

Figure \ref{fig:count_disabled} shows the effect on the number of individuals in the final repertoire when disabling a given input to the controller. The average number of individuals in the repertoire with one input disabled is compared to the repertoire utilizing all the information available. Disabling the nearest neighbor, least frequently visited neighboring square, or the average predicted location results in significant reduction in number of individuals in the repertoire. This is tested using a Rank-Sum statistical test, comparing against repertoires evolved using the full set of inputs.

Another way of quantifying the contribution of the new inputs is to examine the controller parameterization. By reviewing the values of the weights and scales it is possible to infer the contribution of each of the controller inputs to the behavior. Based on this idea it is possible to attribute importance, or some indication of whether each input is used, by slicing the final repertoire along all its 3 dimensions. A slice of an N-dimensional repertoire can be made by locking a value for one of the dimensions and iterating over the remaining ones. The dimension of the locked value is the dimension that is being examined. By extracting all solutions that have this value for the given behavior characteristic, a cross section, or a slice, of the repertoire can be examined. All controllers in this slice have a single behavior characteristic in common, and this allows controllers to be examined in greater detail.

\begin{figure}[h]
\centering
\includegraphics[width=\textwidth]{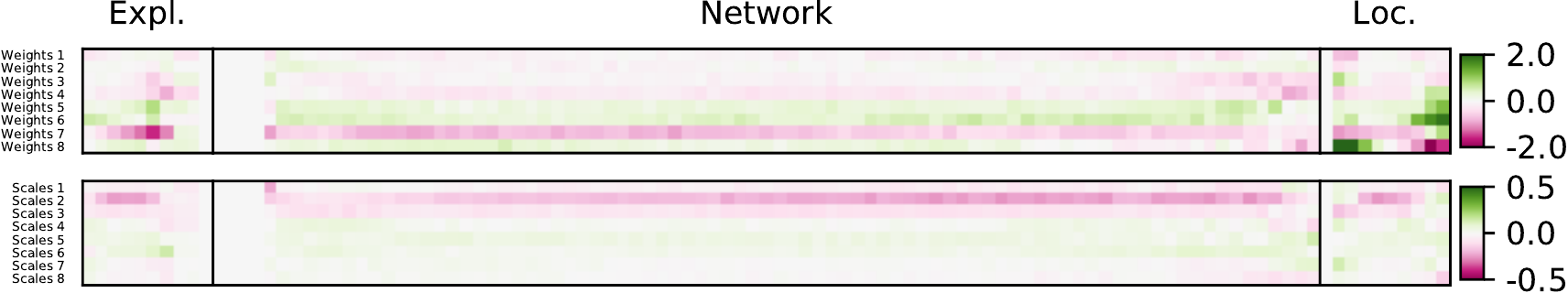}
\centering
\includegraphics[width=\textwidth]{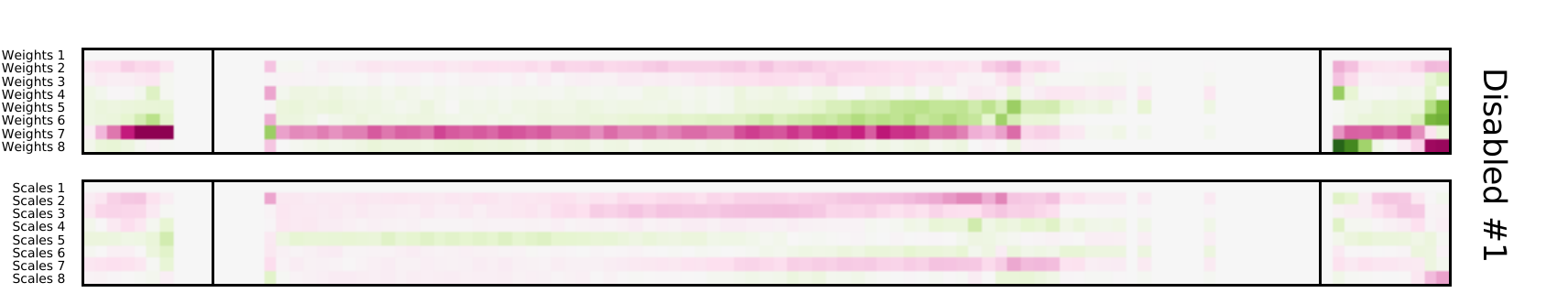}

\includegraphics[width=\textwidth]{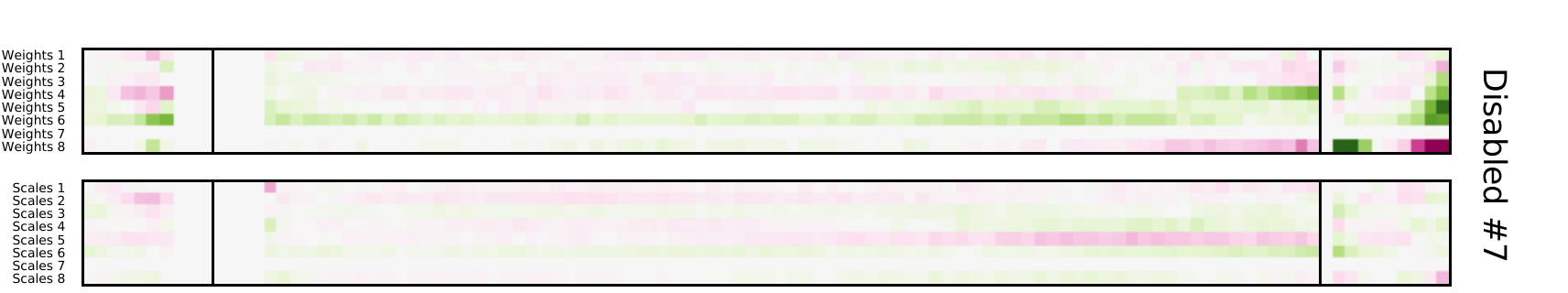}
\includegraphics[width=\textwidth]{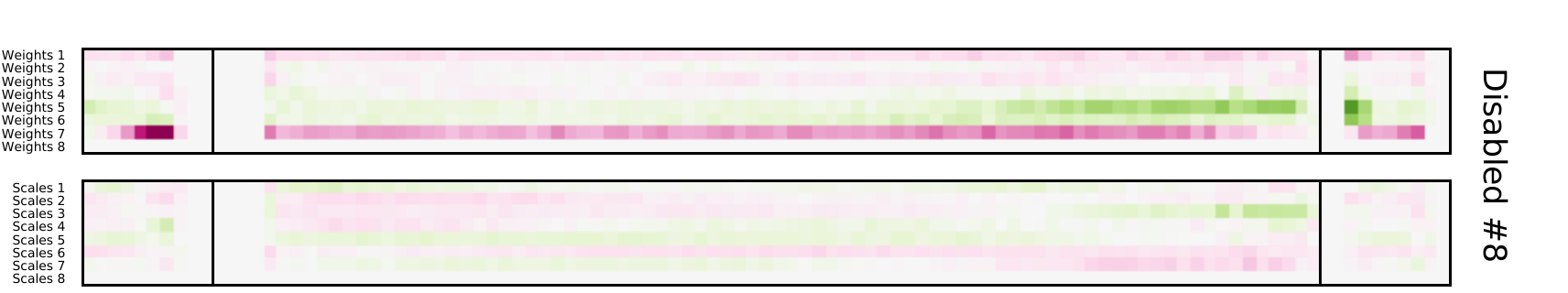}

\caption{Visualization of controller parameter space. Each pair of horizontal subplots are independent of the other pairs. For every slice in a repertoire (a column in this plot), a set of average parameters are calculated. Each pair of horizontal subplots show the average of the average parameters across 8 independent repertoires. The upper pair shows the experiment with all information. The lower three pairs show experiments where input {\#}1, 7 or 8 is disabled. Labels on the left list Weights 1-8 followed by Scales 1-8 for each of the 4 experiments. Each part of the heatmaps from left to right show how each application: exploration, network and localization, require different controller parameters. Each application can be considered independently. The value of the metrics for each application increase from left to right.}

\label{fig:repertoire_analysis}
\end{figure}

Figure \ref{fig:repertoire_analysis} shows a heatmap of the average controller parameters across the solutions in a slice (in any direction) of the repertoire. This type of visualization will be referred to as a parameter heatmap. The resulting output magnitude given the optimized weight and a distance $d_i=100$ is examined. The value of $a_i(100)$ is shown because the sign of the attraction-repulsion component of the Sigmoid-Well function $a_i(d)$ changes around the center point. Showing the value of $a_i(100)$ is a direct way of visualization of whether the weight contributes to a net negative or positive attraction towards the sensed object, without the dependency on the center parameter ($c_i$).

From Figure \ref{fig:repertoire_analysis} it is possible to see that with increasing degree of exploration, Weight {\#}7 is in use and important. This is indicated by the increasing color intensity of the purple color in the top subplot in the row for Weight {\#}7 under the column for exploration. This is sensible, as this particular weight is associated with the least visited neighboring square, and as such, it is intuitive that this weight is used. Another important input for exploration is the use of input {\#}2, which corresponds to the second closest neighbor. The use of a scale parameter indicates that the behaviors are trying to hold a set distance to this neighbor, as opposed to a general attraction or repulsion.

Disabling the closest neighbor (input {\#}1) severely affects the performance in the networking application (Figure \ref{fig:repertoire_analysis}). This can be seen in the parameter heatmap, as a lack of fill (white spaces) in the slices towards the right for the networking application. Without information from the nearest neighbor it becomes very hard to find high performing behaviors, capable of creating and maintaining a lattice structure. In 8 tests with this setup, only a single repertoire had any solutions with high performing networking behaviors. 

The geolocation application relies heavily on two inputs, {\#}7 and {\#}8. These refer to the least visited neighboring square and the average of location predictions. The reliance on the average predicted location is expected, but the need to enable some exploration in order to perform well is unexpected. Behaviors with very high variance in location prediction also seem to require input {\#}5 and {\#}6 (neighbor {\#}5 and {\#}6). This appears to be the result of the evolutionary method discovering that a cluster of agents far away from the average predicated location will lead to high variance in the location predictions.

Disabling the least frequently visited square (Input {\#}7) results in somewhat reduced performance in the exploration application. This requires the geolocation application to rely on other inputs to perform. In particular, the intensity of the color (the magnitude of the parameters) for weight {\#}8 is increased for both high and low variance in geolocation predictions. 

Disabling input {\#}8 (the average predicted position) reduced performance in the geolocation application. Minimum and maximum performance is reduced, and the repertoire is overall somewhat smaller. Of interest is that this also impacts the networking application, as this increases the reliance on input {\#}7 (least visited neighboring square). The compounded effects are of interest. It appears that it is possible to compensate for the loss of some information, given that the controller is optimized with this constraint. 

\section{Discussion}
\label{sec:discussion}

A key motivation in this work was to enable the top-down definition of swarm behaviors and to develop a framework for automating behavior generation. Operators, or even researchers, designing and using swarms are commonly interested in the macroscopic behavior of the swarm, not the low-level interaction between swarm agents. How to find the low-level controllers that enable a given high-level behavior is an unresolved question in swarm research. To this end, this work contributes another method of generating behaviors based on high-level goals or metrics. The methods presented here are powerful, but not complete. It is easy to develop behaviors that are fluid or organic. However, this framework would be less suitable for producing behaviors that require agents to assemble into pre-defined patterns. This is a trade-off, as the controllers in this work were made to be simplistic by design. To enable more complex behaviors may require controllers with an internal state machine, or at the very least, more complex rule-based structures. This would further add to the time required to optimize or evolve the controllers.

Re-evaluating a whole repertoire highlights the issue of combining stochastic metrics with MAP-elites or in general, Quality-Diversity methods. In these experiments, re-evaluation of the entire repertoire resulted in a reduction in repertoire size of up to 76.2\%. Through the use of 5-evaluations per individual this was reduced to 63.0\%, but this is still a drastic reduction in the size of the original repertoire. Multiple evaluations contribute to tackling this challenge. However, as seen from Figure \ref{fig:uncertainty}, there is still room for the behaviors to seemingly move within the repertoire. By doing these experiments it is possible to highlight that noise in Quality-Diversity methods is a challenge. Noise must be considered when designing experiments to discover the true shape of the underlying repertoire. In traditional genetic algorithms it is common to operate with a limited number of elites. In terms of MAP-elites all solutions in the repertoire become an elite and none are re-evaluated. One idea could be to enforce a shelf-life on solutions or require re-evaluation if the solution has persisted in the repertoire for too long. This might remove solutions that get a lucky draw from the a single or a few simulations runs. More research is required to figure out the appropriate measure in order to fully address this issue.

Behavior characteristics can be challenging to design, specifically because evolutionary methods excel at finding ways to exploit metrics without actually providing the intended, or desired, type of behaviors. In this work, exploration is measured by the median visitation count. This was a result of previous experiments using an average metric, which resulted in behaviors merely alternating between two cells instead of actually exploring the area. This provided the same gain in metrics, but did not actually allow for the type of behaviors that were desired. For geolocation, the metric used is the variance of the predicted location. The assumption is that variance decreases as the estimated mean converges on the the true mean. This is often the case but not always. In some very specific cases it is possible to introduce a skew or a bias, where there is a fairly low variance in predictions while most of the predictions are in the wrong place (\cite{engebraaten2015rf}).

The combination of a direct encoding and an open-ended evolutionary method makes it possible to further analyze the results. This would not have been as easy if the controllers had been, for instance, a neural network. Neural networks are inherently hard to fully analyze and understand. In particular, the direct encoded controller made it possible to visualize the effect and contribution of individual controller parameters to the overall swarm behaviors. Understanding the methods in use is key in order to further the field, and having tools that simplifies this is important. This type of analysis would not have been possible with traditional multi-objective optimization, as the intermediate solutions that are not on the Pareto front are discarded. 

The parameter heatmaps might show which controller parameters are in use, but this unfortunately does not paint the entire picture. Disabling the nearest neighbor input resulted in a drastic decrease of the performance in the network application. However, when evolution could use all available inputs the magnitude of Weight {\#}1 and Scale {\#}1 did not indicate that they were heavily used. Likely these parameters are important for performance, but moderate absolute values make them not show up in the parameter heatmaps. While the parameter heatmaps provide a way of visualizing a large multi-dimensional parameter space, there may be parameters that are important for the behavior yet do not have a strong enough contribution to show up clearly in this visualization. In short, the parameter heatmap is another tool to analyze a direct encoded parametric controller. However, it must be used to compliment other methods such as a parameter ablation in order fully understand the evolved swarm behaviors.
\section{Conclusion}
\label{sec:conclusion}

This paper presents a concept for automated behavior generation using evolution for a multi-function swarm. Multi-function swarms have the potential to allow for a new type of multi-tasking previously not seen in swarms. With complex environments and scenarios, it is likely that the operator's needs and requirements will change over time, and as such, the swarm should be capable of adapting to these changes. The viability of evolving large repertoires of behaviors is demonstrated using MAP-elites. These behaviors can be considered behavior primitives that allow for easy adaptation of the swarm to new requirements. It can potentially even be achieved on the fly if simple messages can be broadcast to the entire swarm. This allows the operator to change the behavior based on a change in preferences, desires or other external events. 

Noise is a challenge in MAP-elites. The combination of a greedy algorithm and noisy metrics can result in repertoires that do not reflect the true shape and properties of the underlying system. In this work multiple evaluations is used to reduce the effect of noise. Noise in metrics may enable poorly performing solutions with high variance to outperform better solutions with lower variance due to a lucky draw. Multiple evaluations is not a complete solution, however this study highlights that noise must be considered when applying Quality-Diversity methods such as MAP-elites.

It is possible to investigate the effect each input to a controller has on the swarm performance through parameter ablation. The three most important inputs for this type of artificial physics controller was the nearest neighbor, the least frequently visited neighboring cell and the average predicted emitter location. Results indicate that more information might be better, but more research is required to conclude with certainty. 

Similarly, to the adaptation mechanism presented by \cite{cully2015robots}, it is possible to use a repertoire of behaviors as a way of rapidly adapting to hardware faults or communication errors. In real-world systems communication is unreliable. Having a repertoire could in the future enable even more graceful degradation of performance than what is currently innate within swarms. Optimizing repertoires not only for the three application (exploration, network coverage and localization), but also for varying degrees of allowed communication could bolster the resilience of swarm system. This is future work.



%
%



\bibliographystyle{frontiersinSCNS_ENG_HUMS}
\bibliography{swarm} 
\end{document}